\documentclass[12pt,final]{osuthesis1}

\usepackage{graphicx}

\usepackage{amsfonts}
\usepackage{verbatim}
\usepackage{amsbsy}
\usepackage{amsmath}
\usepackage{amssymb}

\newtheorem{theorem}{\bf Theorem}[section]

\newtheorem{lemma}[theorem]{\bf Lemma}

\title{Evolving Geometries in General relativity} 
\author{Anastasios Taliotis} 
\presentdegrees{MS} 
\unit{Graduate Program in Mathematics}

\gradyear{2010} 

\advisor{Dr. Ulrich Gerlach}

\member{Dr. Andrzej Derdzinski}

\copyrighted   
\copyrightdate{2010} 

\begin{document}
\frontmatter \maketitle

\disscopyright

\begin{abstract}

\hspace{.2in} 
   
The problem of collisions of shockwaves in gravity is well known and has been studied extensively in the literature. Recently, the interest in this area has been revived trough the  anti-de-Sitter space/Conformal Field Theory correspondence (AdS/CFT) with the difference that in this case the background geometry is Anti de Sitter in five dimensions. In a recent project that we have completed in the context of AdS/CFT, we have gained insight in the problem of shockwaves and our goal in this work is to apply the technique we have developed in order to take some farther steps in the direction of shockwaves collisions in ordinary gravity. In the current project, each of the shockwaves correspond to a point-like Stress-Energy tensor that moves with the speed of light while the collision is asymmetric and involves an impact parameter (b). Our method is to expand the metric $(g_{\mu \nu})$ in the background of flat space-time in the presence of the two shockwaves and compute corrections that satisfy causal boundary conditions taking into account  back-reactions of the Stress-Energy tensor of the two point-like particles. Therefore, using Einstein's equations we predict the future of space-time using the fact that we know the past geometry. Our solution respects causality as expected but this casual dependence takes place in an intuitive way. In particular, $g_{\mu \nu}$ at any given point $\vec{r}$ on the transverse plane at fixed $\tau$ evolves according from whether the propagation from the center of each of the shockwaves or from both shockwaves has enough proper time ($\tau$) to reach the point under consideration or not. Simultaneously around the center of each shockwave, the future metric develops a $\delta$-function profile with radius $\tau$; therefore this profile expands outwards from the centers (of the shockwaves) with the speed of light. Finally, we discuss the case of the zero impact parameter collision which results to the violation of conservation and we argue that this might be a signal for the formation of a black hole. 
   
\end{abstract}

\dedication{ \em
I dedicate this to you, May 2010.
}

\begin{acknowl}

\hspace{.2in} 
The author would like to thank Prof.  Andrzej Derdzinski and especially Prof. Gerlach Ulrich for serving in his defense committee and also for stimulating discussions prior and during this thesis. Also the author thanks Prof. Samir Mathur for very informative discussions during the writing of this work. In addition he would like to thank Prof. Thomas Kerler and Herb Clemens for making the transferring from the graduate program of the department of Physics to the graduate program of the department of Mathematics possible and in addition Denise Witcher for guiding him through all the steps of this process. Completion of the graduate courses wouldn't be possible without the help and teaching enthusiasm of Prof. Jean-Francois Lafont, Daniel Shapiro and Joseph Ferrar and particularly Prof. James Cogdell and Alexander Leibman for answering his endless emails with clarity, exactness and always availability. His student mates Corry Christopherson, Fatih Olmez and Zhi Qui have played a crucial role during the coursework as not only they kept encouraging him but also have spent enormous time from their time in order to patiently answer all of his questions. Attending many of the required classes would had been impossible without the help of his best friend Chen Zang who baby sat the authors son, Nikolas Alexandrou Taliotis, while the author and his wife Maria Alexandrou had to attend classes. The author owns special gratefulness to his physics advisor, Prof. Yuri Kovchegov, for teaching him physics, for teaching him how to see through the complicate mathematics and nail down the physical picture and especially for showing him how to undertake his responsibilities not only inside but also outside of the academia. Special thanks to Prof. Tom Banks, Steve Giddings, Yuri Kovchegov and Krishna Rajagopal for their encouragement in submitting this thesis on the arXiv after it was defended. Lastly but most important, the author would like to thank Maria Alexandrou for all of her patience during these years and for lifting most of the weights of their journey in life while keep smiling and being an example of a kind human, a reliable partner and a wonderful woman: he delicates this work to Her.

 This work is sponsored in part by the U.S. Department of Energy under Grant No. DE-FG02-05ER41377 and in part by the Institution of Governmental Scholarships of Cyprus (IKY).

\end{acknowl}

\begin{vita}
  \begin{datelist}
  	\item[1977] Born in Nicosia, Cyprus 
  	\item[2003] B.Sc. in Physics,  
	\item[2008]  MS in Physics

  \item[2008-Present] Graduate Research Associate in the Department of Physics and Graduate Student in the Department of Mathematics of The Ohio State University
  \end{datelist}

 \begin{publist}

\pubitem{ A.~Taliotis,
  ``DIS from the AdS/CFT correspondence,''
  Nucl.\ Phys.\  A {\bf 830}, 299C (2009)
  [arXiv:0907.4204 [hep-th]].
}

\pubitem{ J.~L.~Albacete, Y.~V.~Kovchegov and A.~Taliotis,
  ``Heavy Quark Potential at Finite Temperature in AdS/CFT Revisited,''
  Phys.\ Rev.\  D {\bf 78}, 115007 (2008)
  [arXiv:0807.4747 [hep-th]].
}

\pubitem{J.~L.~Albacete, Y.~V.~Kovchegov and A.~Taliotis,
  ``Asymmetric Collision of Two Shock Waves in AdS$_5$,''
  JHEP {\bf 0905}, 060 (2009)
  [arXiv:0902.3046 [hep-th]].}

\pubitem{J.~L.~Albacete, Y.~V.~Kovchegov and A.~Taliotis,
  ``Modeling Heavy Ion Collisions in AdS/CFT,''
  JHEP {\bf 0807}, 100 (2008)
  [arXiv:0805.2927 [hep-th]].}

\pubitem{ J.~L.~Albacete, Y.~V.~Kovchegov and A.~Taliotis,
  ``DIS on a Large Nucleus in AdS/CFT,''
  JHEP {\bf 0807}, 074 (2008)
  [arXiv:0806.1484 [hep-th]].
}

\pubitem{ Y.~V.~Kovchegov and A.~Taliotis,
  Phys.\ Rev.\  C {\bf 76}, 014905 (2007)
  [arXiv:0705.1234 [hep-ph]].
 }

 \end{publist}

\begin{fieldsstudy}
    \majorfield{Mathematics}
    \specialization{High Energy Physics and Applied Mathematics}
  \end{fieldsstudy}
\end{vita}

\tableofcontents
\listoffigures

\mainmatter

\everymath{\displaystyle}

\chapter{Introduction} \label{Intr} 


The problem of the collision of two shoockwaves in four dimensions (with Lorentz signature) that result from boosting two black holes to the speed of light  is well known and has been extensively studied in the literature. References  \cite{DEath:1992hb,DEath:1992hd,DEath:1992qu,Aichelburg:1970dh,Dray:1984ha,Hotta:1992qy,Hawking:1969sw,Sfetsos:1994xa,Eardley:2002re} provide only a subset of the work dedicated to the particular problem (for some interesting recent developments see \cite{Giddings:2009gj,Giddings:2010pp}). In particular, \cite{DEath:1992hb,DEath:1992hd,DEath:1992qu} deals with this problem pertubatively and in a series of three papers the authors compute the metric and derive a formula for the gravitational radiation. The collision is axisymmetric while the bulk matter that creates the shockwaves as well as the back-reaction effects are not taken into account.

Recently, the high energy physics  community has also expressed special interest in the topic of shockwaves collisions in the framework of General Relativity \cite{Taliotis:2010pi,Albacete:2009ji,Bartels:2009bv,Albacete:2008vs,Grumiller:2008va,Gubser:2009sx,Gubser:2008pc,Kovchegov:2009du,DuenasVidal:2010vi}
through the anti-de-Sitter space/Conformal Field Theory correspondence (AdS/CFT) \cite{Maldacena:1997re,Witten:1998qj,Gubser:1998bc,Gubser:1997yh,de Haro:2000xn}.  The AdS/CFT duality, allows to formulate a process that is associated with non-abelian gauge theories at strong coupling as a purely gravitational problem. As a result, one may map the problem of heavy ion collisions (in four dimensions) onto shockwaves collisions in gravity in five dimensions. Recently, we have been involved in such problems \cite{Taliotis:2010pi,Albacete:2009ji,Bartels:2009bv,Albacete:2008vs} in gauge theories at strong coupling whose dual five dimensional gravitational description, exhibits similar characteristics with those of shockwave collisions in (ordinary) four dimensional gravity.

Our goal in this work is to attempt to apply our earlier experience \cite{Taliotis:2010pi,Albacete:2009ji,Bartels:2009bv,Albacete:2008vs} and especially apply the technique that we have developed in \cite{Taliotis:2010pi} in (ordinary gravity in) four dimensions. We choose to work in a different coordinate system than \cite{DEath:1992hb,DEath:1992hd,DEath:1992qu} \footnote{More precisely our coordinate system coincides with the initial coordinate system of \cite{DEath:1992hb,DEath:1992hd,DEath:1992qu}. However, in the process of the calculation \cite{DEath:1992hb,DEath:1992hd,DEath:1992qu} choose to change coordinates.} in order to retain the geometrical insight of the collision and in addition we take into account the matter responsible for the creation of the  shockwaves and the back-reaction effects as well. Furthermore, the collision we consider here is not axisymmetric but is involves a non zero impact parameter.

\vspace{0.3in}
We organize the paper as follows.

\vspace{0.3in}

In chapter \ref{Sup} we state the problem we want to solve and construct the main set up. Our goal is  to determine the evolution of the geometry assuming that we know it in some time interval (negative times). The initial geometry is given by two shockwaves which correspond to  a non zero Stress-Energy tensor \footnote{This Stress-Energy tensor is due to two point-like particles moving opposite to each other (see figure \ref{BTen}) with the speed of light and begin to interact for positive times}. Our method  is to construct a perturbative approach by expanding the metric around the background given by the flat metric along with the two shockwaves. Equation (\ref{s12}) shows the form of the metric at all times while figure \ref{interaction} offers a diagrammatical intuition of the terms of the metric we attempt to calculate in this project.

In chapter \ref{B2B} we take into account the interaction of the one particle with the gravitational field created from the other and vice versa. In terms of Feynman diagrams, loosely speaking, these corrections correspond to the diagrams of figure \ref{SelfInt4d}. The corrections to the Stress-Energy tensor corresponding to these diagrams along with the corrections of the metric tensor corresponding to diagram of figure \ref{interaction} form a consistent set \footnote{By consistent we mean with the order of the parameter we are expanding (see also (\ref{Rmn2gt})).} of the  corrections that have to been taken into account. We verify that the modified Stress-Energy tensor is conserved (to the order of the expansion that we are working) and we find that it is traceless. This last condition results into some pleasing simplifications for Einstein's equations (compare (\ref{Ein}) with (\ref{EE})).

Chapter \ref{FE} deals with the field equations and the specification of the gauge. Our attempt is to perform a perturbative calculation about a metric that looks almost flat but contains two shockwaves moving opposite to each other and colliding (see (\ref{s12})). The shockwaves provide an effective Stress-Energy tensor in addition to the (actual) Stress-Energy tensor (see (\ref{Rmn2gtT})) that creates the two shockwaves; these terms correspond to products of the form $t_{1}t_{2}$ and $T_{\mu \nu}$ of equation (\ref{deq1}) respectively. A suitable gauge choice simplifies the field equations (\ref{deq1}) to equations (\ref{deq}) which are solved in the next chapter.

In Chapter \ref{SEq} we specify the boundary conditions of the field equations (\ref{deq}) and the corresponding Green's function. In particular, we seek for causal solutions and therefore the associated Green's function to the differential operator (\ref{box}) is the retarded Green's function. We show how the integrations on the light-cone and transverse plane may be performed omitting some of the intermediate steps for Appendices \ref{A}, \ref{B} and \ref{C}. Eventually we derive a formula for the metric tensor, equation (\ref{gmn2}), to the order we are working. This is our final result and it generally has the structure (\ref{GMN}).

Finally in Chapter \ref{Conc} we discuss about the area of validity of our solution and summarize our conclusions. In particular, we argue that the presence of matter and the back-reaction effects may not be ignored as they result to an important contribution to the metric. We also see that as the impact parameter tends to zero, the metric diverges logarithmically. It is in our belief that this is a signal that a classical approach to the problem stops being valid and that a quantum description is required. Lastly, we talk about the general form of the metric. Although its evolution is constrained by casualty as expected; this evolution takes place in an intuitive way: At a given (proper) time, any arbitrary point on the transverse plane evolves according from whether the signal from the center of the one or the other shockwave or both, has enough time to reach the point under consideration. We had encountered such a behavior in an analogous set up in \cite{Taliotis:2010pi} although the geometry there was Anti de Sitter geometry in five dimensions. In \cite{Taliotis:2010pi} we had claimed that a similar evolution of the metric should also be observed in four dimensions; in this work we verify our conjecture.

\chapter{Setting up the problem} \label{Sup} 

\section{Single Shockwave Solution}\label{1sw}

We begin by defining the coordinate system (gauge) we work. We choose to work in light-cone coordinates defined by

\begin{equation}\label{LC}
x^{\mu}=(x^+,x^-,x^1,x^2) \hspace{0.3in}x^{\pm}=\frac{x^0\pm x^3}{\sqrt{2}}
\end{equation}
where $x^0$ is the time axis and $x^1,x^2,x^3$ cover $\mathbb{R}^{3}$. The convection for the flat metric that we use is
\begin{equation}\label{nmn}
g_{\mu \nu}=(-1,1,1,1)
\end{equation}

We suppose that we have a black hole metric that we boost to the speed of light along a given direction ($x^3$ direction). The metric is known \cite{Aichelburg:1970dh} and is given by

\begin{equation}\label{ds1}
ds^2 \, = g_{\mu \nu}dx^{\mu } dx^{\nu}= -2 \, dx^+ \, dx^- + t_1
(x^+,x^1,x^2)  d x^{+ \, 2} + d x_\perp^2 , \hspace{0.3in}  
\end{equation}
where
\begin{align}\label{s1}
t_1=-\mu \delta(x^+) \log(k r) \hspace{0.3in}\vec{r}=(x^1,x^2) \hspace{0.3in}  r=\sqrt{(x^1)^2+(x^2)^2}.
\end{align}
According to equation (\ref{ds1}) we denote the transverse flat metric by $d x_\perp^2 = (d x^1 )^2 + (d x^2)^2$ while $\delta(x^+)$ denotes the delta (Dirac) function. The parameter $\mu$ has dimensions of length and its physical meaning will become apparent in what follows while $k$ serves as an ultraviolet cutoff and whose physical meaning is discussed in the conclusions (see Section (\ref{scon})).

One may check directly whether (\ref{ds1}) solves Einstein's equations \footnote{The cosmological constant is assumed zero.} which may be cast as

\begin{align}\label{Ein}
R_{\mu \nu}  = \kappa_4^2
\left( T_{\mu \nu} - \frac{1}{2} \, g_{\mu \nu} \, T \right) \hspace{0.3in}T = T_{\mu}^{\mu} = \, T_{\mu \nu} \, g^{\mu \nu}\hspace{0.35in}\kappa_4^2=8\pi G_4
\end{align}
where $R^{\mu \nu}$ is the Ricci tensor, $T_{\mu \nu}$ the Stress-Energy tensor and $G_4$ the Newton's constant (in four dimensions).

Direct substitution of (\ref{ds1}) in the formula that computes $R_{\mu \nu}$ results to

\begin{equation}\label{Rmn}
R_{\mu \nu}=\delta_{\mu +} \delta_{\nu +}\left(\frac{2 \pi}{2}\mu \delta(x^+)\delta^{(2)}(\vec{r})  \right)
\end{equation}
where $ \delta_{\nu +}$ is a Kronecker delta. This implies that all components of $R_{\mu \nu}$ are zero except from $R_{++}$. In order to arrive to equation (\ref{Rmn}) we had to evaluate the following linear differential expression
\begin{equation}\label{R++log}
R_{++}=-\frac{1}{2}\nabla_{\perp}^2 t_1(x^+,x^1,x^2).
\end{equation}
where $\nabla_{\perp}^2$ is the Laplace operator in two dimensions while we used the identity
\begin{equation}\label{dlog}
\nabla_{\perp}^2 \log (k r)=2 \pi\delta^{(2)}(\vec{r}).
\end{equation}

Equations (\ref{Ein}) and (\ref{Rmn}) imply that the metric tensor of equation (\ref{ds1}) corresponds to a Stress-Energy tensor. Shifting the origin \footnote{The reason for this shift will become apparent in the next section.} along negative $x^1$ for distance $b$  we find that the Stress-Energy tensor is given by 

\begin{equation}\label{T1}
T_{\mu \nu}^{(1)}=\delta_{\mu +} \delta_{\nu +}\left(\frac{\pi\mu}{\kappa_4^2}\delta(x^+)\delta(x^1-b)\delta(x^2) \right).
\end{equation}
The presence of the superscript $(^{(1)})$ on $T_{++}$ of (\ref{T1}) is to highlight that it is of first order in the parameter
$\mu$ \footnote{Generally, the superscripts emphasize the number of times the source $t_i$ (i=1,2 assuming we have two sources) appears. Generally in the object $A^{(n)}$ there exists the product  $t_1^k t_2^{n-k}$ or any linear combination of differentiation/integration of $t_1$ and $t_2$ with respect to their arguments $x^{\mu}$.}. This equation implies that the shockwave of (\ref{ds1}) is a consequence of a point particle moving along the $x^-$ direction with the speed of light and hence its massless. Indeed, the ratio $\frac{\mu}{\kappa_4^2}$ has dimensions of mass as should. One may check that $T_{\mu \nu}$ of (\ref{T1}) is covariantly conserved. 

\begin{figure}
\centering
\includegraphics[scale=0.5]{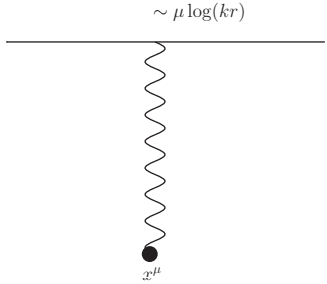}
\caption{The shockwave solution: a graviton (curly line) is emitted from the bulk source (straight line) with coupling $\mu \log(kr)$ which is measured at point $x^{\mu}$. This is a very special case where a single graviton exchange between the source and the bulk happens to be an exact solution to the non linear Einstein's equations.}
\label{vertex}
\end{figure}

We may gain some insight in the physical system at hand if we use a diagrammatical approach. Despite that General Relativity is a non linear theory, the metric (\ref{ds1}) satisfies a linear differential equation. The fact that (\ref{ds1}) is an expression of a perturbation of a flat metric proportional to $\mu$ suggests the diagram of figure \ref{vertex}. 
It represents the measurement of the gravitational field at point $x^\mu$, which, loosely speaking, is created by a single graviton emission from the source (point-like Stress Energy tensor) of equation (\ref{T1}) with effective coupling proportional to $\mu \log(k r)$.  In other words for this special case of a single shockwave, the first order solution happens to be the exact solution to all orders.

\section{Superimposing two Shockwaves}\label{2sw}

\begin{figure}
\centering
\includegraphics[scale=0.5]{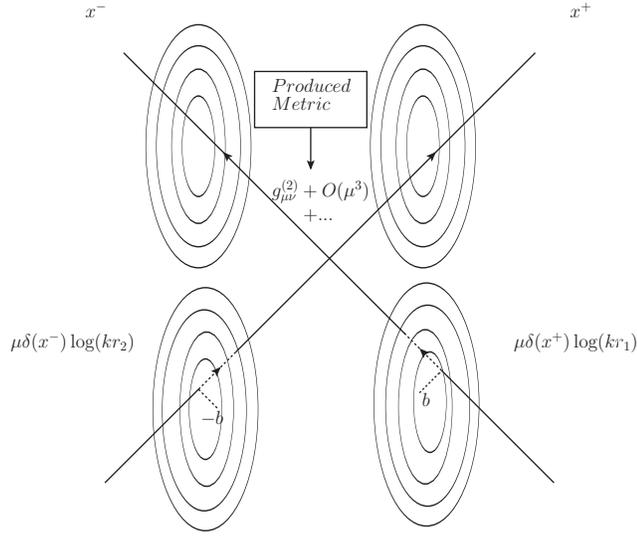}
 \caption{The two shockwaves before and after the collision moving along $x^{\pm}$ axis and dragging a perpendicular gravitational field which is constant along the circular lines. They collide at the origin producing a gravitational field in the forward light cone. Our goal to compute the ``produced" metric and in particular $g_{\mu \nu}^{(2)}$.}
  \label{offcenter}
 \end{figure}

Having defined all the necessary ingredients we now proceed to the main part of the setup. We want to superimpose two such shockwaves whose sources are two point-like distributions  of matter moving towards each other \footnote{This is true for negative times only. For positive times their trajectories are altered and have to be specified taking into account back-reaction effects (see Chapter \ref{B2B}).} in space-time. We want to collide these shockwaves (and as a result the corresponding Stress-Energy tensors as well) at a non-zero impact parameter and hence study the problem within the classical theory of gravity. Therefore, $T_{\mu \nu}$ has in addition to (\ref{T1}) the symmetric part 

\begin{align}\label{T2}
T^{(1)}_{--} \, = \frac{\pi \mu}{\kappa_4^2} \delta (x^-) \, \delta (x^1+b)\delta (x^2)
\end{align}
which creates a second shockwave. In terms of space-time, this would correspond in ``colliding" two metrics in an off center process. Figure \ref{offcenter} represents the four dimensional picture, right before the collision of the two shockwaves. Following \cite{Albacete:2009ji},\cite{Albacete:2008vs}, the metric that describes the process should look like

\begin{figure}
\centering
\includegraphics[scale=0.5]{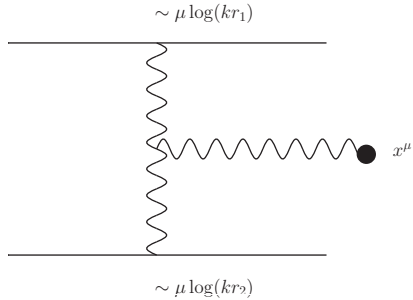}
\caption{The Feynman diagram that represents the $\mu^2$ correction of the metric:  It represents along with the diagram of figure \ref{SelfInt4d} (see Chapter \ref{B2B}), the first non trivial correction to (\ref{s12}). It shows how the two metrics that each one looks like (\ref{s1}) merge. The gravitational field is measured at the point $x^{\mu}$.}
\label{interaction}
\end{figure}

\begin{align}\label{s12}
ds^2 \, &= -2 \, dx^+ \, dx^- + d x_\perp^2 + t^{(1)}_1(x^+,x^1-b,x^2)  \, d x^{+ \, 2} \notag\\&
 + t^{(1)}_2(x^-,x^1+b,x^2)  \, d x^{- \, 2}
+ \theta(x^+)\theta(x^-)g^{(2)} _{\mu \nu}(x^{\kappa},z)dx^{\mu}dx^{\nu}   + \ldots    , \notag\\&
 t^{(1)}_{1,2}(x^1 \mp b,x^2)= -\mu \log\left (k \sqrt{(x^1 \mp b)^2+(x^2)^2}\right) \delta(x^{\pm}).
\end{align}
The first three terms correspond to the flat (Minkowski) space. The next two are of first order in $\mu$ and are created by the two point-like particles. These move (initially) towards each other along $x^3$ and they have an impact parameter $2b$ along the $x^1$ axis as figure \ref{BTen} depicts.  As they are the sources of the two shockwaves, they correspond to two vertex diagrams that look like the one in figure \ref{vertex}. This is a superposition of two metrics with each one looking like (\ref{s1}). However, the non-linearities of the gravitational field require higher order terms. The second order corrections are explicitly displayed in (\ref{s12}) and they appear once the two shockwaves cross each other; in the forward light cone. This is precisely the meaning of the $\theta$-functions; they emphasize that the metric (\ref{s12}) solves Einstein's equations  exactly in the presence of both shockwaves only for negative $x^{\pm}$. The additional terms of the metric appear in the forward light cone only and describe the effects of collision. The main work of this paper is to show how these terms may be calculated to order $\mu^2$, that is find $g_{\mu \nu}^{(2)}$. The second order correction in $\mu$ of $g_{\mu \nu}$ corresponds to the diagram of figure \ref{interaction}.
\chapter{Back-Reactions} \label{B2B} 

\section{Corrections to $T_{\mu \nu}$ and Geodesics}\label{Tgen}

As has been already mentioned below (\ref{T1}), $T_{++}^{(1)}$ is conserved in the gravitational field of (\ref{ds1}), (\ref{s1}). In fact, conservation for this case happens to be valid to all orders in $\mu$ \footnote{In practice, only the first order in $\mu$ appears in the resulting equations. This is in accordance with our intuitive picture of figure \ref{vertex}: Gravity behaves linearly with respect to the metric (\ref{ds1}).}
\begin{align}\label{con1}
\nabla^{M} J_{\mu \nu}^{(1)}=0.
\end{align}

\begin{figure}
\centering
\includegraphics[scale=0.7]{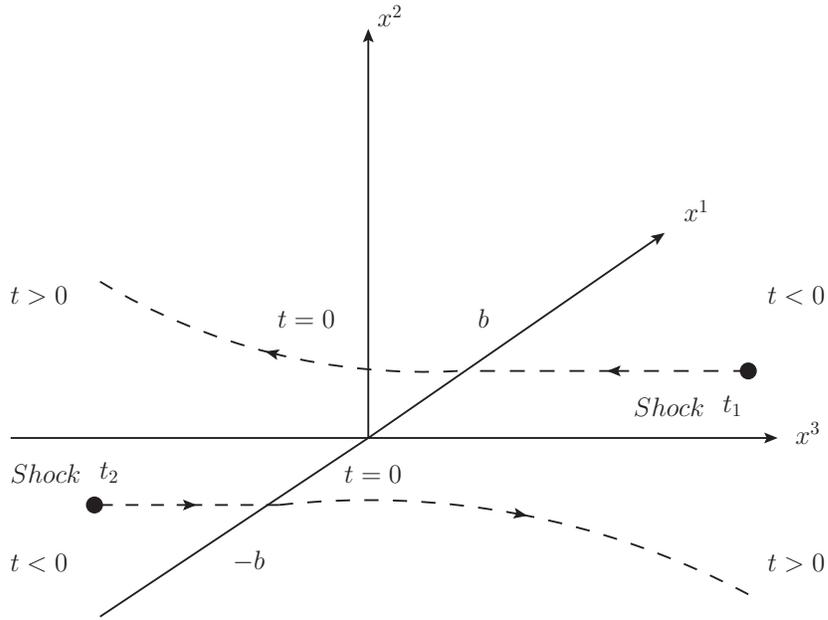}
\caption{The sources (point-like particles) represented as black dots and their trajectories:  For negative times they move along straight lines.  At t=0 each particle (dot) intersects the shock due to the other particle and its trajectory suffers a kick (see curved path); hence $T_{\mu \nu}$ changes with time.}
\label{BTen}
\end{figure}
\begin{figure}
\centering
\includegraphics[scale=0.8]{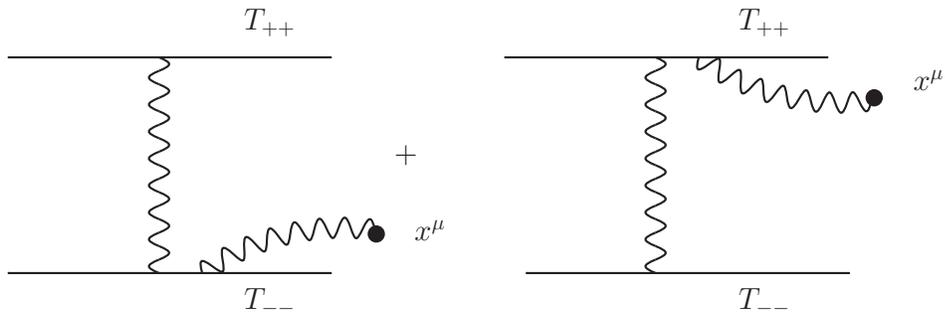}
\caption{Graviton emission diagrams resulting from self corrections to $T_{\mu \nu}$. The first source interacts via the gravitational field created by the other and vice versa. The point $x^\mu$ is the space-time point where $g_{\mu \nu}$ is measured. }
\label{SelfInt4d}
\end{figure}

Conservation to first order is still valid when we consider simultaneously $T_{++}^{(1)}$ and $T_{--}^{(1)}$ in the presence of the gravitational field (\ref{s12}). However, this is no longer true at the second order in $\mu$. The reason is because the $T_{--}$ ( $T_{++}$) source moves in the gravitational field of the $t_1$ $(t_2)$ shockwave, altering its initial trajectory. Figure \ref{BTen} outlines what happens while figure \ref{SelfInt4d} offers a diagramatical intuition regarding the self-corrections to $T_{\mu \nu}$. This implies that we should correct $T_{\mu \nu}^{(1)}$ in order to preserve conservation (of the total Stress-Energy tensor). However, since we do not know the nature (equation of state) of $T_{\mu \nu}$ we make the assumption that these objects interact only via gravitational forces \footnote{More precisely, we assume that any other interactions are small compared to the gravitational forces.}.

\section{Calculating the Corrections for $T_{\mu \nu}$}\label{Tcor}
Since these particles are point-like and massless, they should travel along null geodesics; as it is rigorously shown in \cite{Papa:1974} conservation of (the total) $T_{\mu \nu }$ is then guaranteed. This suggests that we need

\begin{align}\label{Tpp}
T ^{\mu \nu}= \frac{\pi \mu}{\kappa_4^2}  \sum_{(I=1)}^2   \dot{x_{(I)}^{\mu}}\dot{x_{(I)}^{\nu}}  \frac{1} {\sqrt{-g}}\delta^{(3)} \left(\vec{x}_{(I)}- \vec{x}_{(I)}(s_{(I)}) \right)         
\end{align}
which gives the total Stress -Energy tensor of both point-like particles of mass $\frac{\pi \mu}{\kappa_4^2}$ each moving along the trajectory $ \vec{x}_{(I)}(s_{(I)})$ parameterized by $s_{(I)}$. The quantity $g$ is the determinant of the (total) metric tensor, the factor $\pi$ is in agreement with our convention (that reproduces (\ref{T1}) - see below) while the dots denote differentiation with respect to the parameter $s_I$. Before calculating higher order corrections to $T_{\mu \nu}$, we find it instructive to check whether this formula reproduces (\ref{T1}) in the case of one particle ($I=1$). The trajectory of this particle which moves ultra-relativistic ally along negative $x^3$ is parameterized by

 \begin{align}\label{tr1c}
x^{\mu}_{(1)}(x^0,x^1,x^2,x^3)=(t,0,0,-t).
\end{align}
In light cone coordinates and choosing to parameterize the trajectory by $x^-$, equation (\ref{tr1c}) then implies
 \begin{align}\label{tr1lc}
x^{\mu}_{(1)}=(x^+(x^-)=0,x^-(x^-);x^1(x^-)=0,x^2(x^-)=0)=(0,x^-;0,0).
\end{align}
Direct substitution of (\ref{tr1lc}) to (\ref{Tpp}) yields
\begin{align}\label{Tp1}
T ^{--}_{(1)}=T_{(1)++}= \frac{\pi \mu}{\kappa_4^2} \delta(x^+) \delta(x^1)\delta(x^2) 
\end{align}
which is exactly equation (\ref{T1}) (for $ b=0$). This computation also clarifies the convection of $\pi$ in (\ref{Tpp}). The part of $T_{\mu \nu}$ of equation (\ref{T2}) due to the second particle is reproduced similarly.

The next step is to calculate the next (second) order corrections (in $\mu$) of $T_{\mu \nu}$ which translates into finding the corrections to the trajectories $x_{(I)}$. These may be obtained from the geodesic equations. In particular, we only need the first order corrections to $x_{(I)}$
as we already have a power of $\mu$ in front of the summation operator (see (\ref{Tpp})). The geodesic equations we need read

\begin{align}\label{geo}
\ddot{x^{\mu}}_I+\Gamma_{J;\nu\rho}^{\mu}  \dot{x_I^{\nu}}\dot{x_I^{\rho}}=0 \hspace{0.3in}I,J=1,2 \hspace{0.3in}I\neq J
\end{align}
and are interpreted as the motion of particle $I$ in the gravitational field of the particle $J$  (due to $\Gamma_{J;\nu\rho}^{\mu}$ where $\Gamma$ are the Christoffel symbols) and vice versa; this is precisely the meaning of the subscripts $I$ and $J$.
We begin with computing the corrections to the particle $I=1$ whose (first order) perturbed trajectory looks like \footnote{We drop the subscript $_{(1)}$ which labels the particle $I=1$ for simplicity.}
 \begin{align}\label{tr11}
(x^{\mu}(x^-))^{(0)}+(x^{\mu}(x^-))^{(1)} =  \left((x^+)^{(1)},x^- + (x^-)^{(1)},b+(x^1)^{(1)},(x^2)^{(1)} \right)
\end{align}
where we have chosen to parameterize the trajectory with $x^-$ (i.e. $s_{(I=1)}=x^-$). The superscript $^{(1)}$ denotes the order in the expansion. Taking into account (\ref{Tpp}) and the fact that $\Gamma_{(2)\nu\rho}^{\mu} \sim \mu$ otherwise is zero, we deduce that both of the terms $\dot{x^{\nu}}$ or $\dot{x^{\rho}}$ have to be of zeroth order; i.e. the only choice is $\nu=\rho=-$. This implies that we need to determine $\Gamma_{--}^{\mu}$ to first order in $\mu$ that arise from the second particle \footnote{Where we (also) dropped the subscript $_{(2)}$ which labels the second particle.} and which read
\begin{align}\label{Gam}
\Gamma_{--}^{+}=  -  \frac{1}{2} t_{2,x^-}     \hspace{0.3in}     \Gamma_{--}^{-}=  0  \hspace{0.3in}\Gamma_{--}^{1}=  -  \frac{1}{2} t_{2,x^1}   \hspace{0.3in}\Gamma_{--}^{2}= -   \frac{1}{2} t_{2,x^2} .
\end{align}
A few explanations about our notation are in order: the term $t_2$ is due to the second particle and is given in equation (\ref{s12}). The subscript $_{2,x^{\mu}}$ on $t_2$ denotes ordinary differentiation of the source \footnote{As matter implies curvature and curvature (a non-flat metric) implies matter, we will often use these notions interchangeably while their distinction should be evident from the context. } $t_2$ with respect to the coordinate $x^{\mu}$ . According to (\ref{s12}), the source $t_2$ is of first order in $\mu$ \footnote{From now on we drop the superscript $^{(1)}$ which denotes the order in $\mu$ from $t_{1,2}^{(1)}$.} and as a result the same applies for the $\Gamma$'s of (\ref{Gam}); it should be by now obvious that the Christoffel symbols are due to the second particle ($t_2$).

The final step is to integrate (\ref{geo}) using (\ref{Gam}) and using causal boundary conditions and substitute the result in formula (\ref{Tpp}). As we are interested in second order corrections in $\mu$, we immediately conclude that at least one of  $\dot{x^{\mu}}$ or $\dot{x^{\nu}}$ should be of order zero, i.e. $\mu=-$ or $\nu \neq-$ or $\mu \neq-$, $\nu =-$. We also note that $\sqrt{-g}\sim 1+O(\mu^2)$ and hence according to (\ref{Tpp}) corrections from $\sqrt{-g}$ do not contribute to $T_{\mu \nu}$ at $O(\mu^2)$. We consider two cases

\vspace{0.3in}
\underline{Case I:  $\sqrt{-g}=1$, $\mu=\nu=-$}
\vspace{0.3in}

In this case the modification of $T_{\mu \nu}$ of equation (\ref{Tpp}) to this order we are working is in the arguments of the delta's. We have \footnote{ All the integrations with respect to $dx^{\pm}$ that follow from now on (see also Appendices \ref{A} and \ref{B}) will imply the obvious: $\int dx^-$ stands for $\int_{-\infty}^{x^-} dx'^-$, $\int dx^- \int dx^-$ stands for $\int_{-\infty}^{x^-}dx'^-\int_{-\infty}^{x'^-} dx''^-$ etc.}
\begin{align}\label{TI1}
T ^{--}= &\frac{\pi \mu}{\kappa_4^2}   \delta \left(x^+ -   \frac{1}{2}\int dx^-  \int  dx^- t_{2,x^-} \right)              \delta \left(x^1-b -\frac{1}{2}  \int dx^-\int dx^- t_{2,x^1} \right)  \notag\\&           
 \hspace{0.7in} \times
 \delta \left(x^2 - \frac{1}{2}\int dx^-\int dx^- t_{2,x^2} \right)              
\end{align}
Expanding the delta's to first order in the sources we obtain
\begin{align}\label{TI2}
T ^{--}=& \frac{\pi \mu}{\kappa_4^2} \Bigg[ \delta(x^+)\delta(x^1-b)\delta(x^2)-\frac{1}{2}\int  dx^- \Bigg( t_{2} \hspace{0.02in} \delta'(x+)\delta(x^1-b)\delta(x^2) 
\notag\\&
+\int dx^- \Big( t_{2,x^1}\delta(x^+)\delta'(x^1-b)\delta(x^2)+ t_{2,x^2}\delta(x^+)\delta(x^1-b)\delta'(x^2) \Big)\Bigg) \Bigg].
\end{align}
We may cast (the first order correction terms of the) last equation in a compact form by expressing it in terms of $t_1$ and $t_2$. Using the identity
\begin{align}\label{nab1}
\mu \delta(x^{\pm}) \delta^{(2)}(\vec{r} - \vec{b}_{1,2})=-\frac{1}{2 \pi} \nabla_{\bot}^2 t_{1,2}\hspace{0.3in}\vec{b}_{1,2}=(\pm b,0)
\end{align} 
(see (\ref{dlog}) and (\ref{s12})) the $O(\mu^2)$ terms of (\ref{TI2}) take the form

\begin{align}\label{TI3}
(T_{(1)++})^{(2)}&=(T_{(1)} ^{--})^{(2)}= \frac{1}{4\kappa_4^2} \notag\\&
 \times \int  dx^-\left(  t_{2} \nabla_{\bot}^2 t_{1,x^+}  +\nabla_{\bot}^2 t_{1,x^1} \int dx^- t_{2,x^1}+\nabla_{\bot}^2 t_{1,x^2} \int dx^- t_{2,x^2}\right)
\end{align}
where we restored the subscript $_{(1)}$ and the superscript $^{(2)}$ in order to highlight that this is the second order correction to $T_{\mu \nu}$ of the first particle.

\vspace{0.3in}
\underline{Case II:  $\sqrt{-g}=1$, $\mu=-$, $\nu \neq -$}
\vspace{0.3in}

In this case the modification of $T_{\mu \nu}$ of equation (\ref{Tpp}) to this order we are working is in the factor $\dot{x}^{\nu}\dot{x}^-=\dot{x}^{\nu}$. Combining (\ref{geo}) and (\ref{Gam}) one may compute $\dot{x}^{\nu}$. Plugging this result to (\ref{Tpp}) and employing the identity (\ref{nab1}) in order to write the transverse delta's in terms of $t_1$ yields to
\begin{subequations}\label{TpII}
\begin{align}
&  (T^{+-}_{(1)})^{(2)}=  -   \frac{1}{4\kappa_4^2}t_{2}                  \nabla_{\bot}^2 t_{1} \label{TpII+-}\\&
    (T_{(1)+1})^{(2)}=-(T^{-1})^{(2)}=              \frac{1}{4\kappa_4^2} \nabla_{\bot}^2 t_{1} \int  dx^- t_{2,x^1}        \label{TpII+1} \\ &
    (T_{(1)+2})^{(2)}=-(T^{-2})^{(2)} = 	        \frac{1}{4\kappa_4^2} \nabla_{\bot}^2 t_{1} \int  dx^- t_{2,x^2}. \label{TpII+2}
\end{align}
\end{subequations}

\underline{Second order corrections to the total $T_{\mu \nu}$}
\vspace{0.15in}

The second order corrections to the stress energy tensor $(T_{(1) \mu \nu})^{(2)}$ of the first particle is given by equations (\ref{TI3}) and (\ref{TpII}). The corrections $(T_{(2) \mu \nu})^{(2)}$ of the second particle may be found analogously and therefore the second order corrections to the total stress energy tensor read 
\begin{subequations}\label{Tmn2}
 \begin{align}
&  (T_{+-})^{(2)}= -  (T^{+-})^{(2)}=    \frac{1}{4}\frac{ 1}{ \kappa_4^2}     \left(   t_{2} \nabla_{\bot}^2 t_{1} + t_{1} \nabla_{\bot}^2 t_{2} \right) \label{T+-}   \\&
(T_{++})^{(2)}= \frac{1}{4}\frac{ 1}{\kappa_4^2} \int  dx^-\left(  t_{2}  \nabla_{\bot}^2 t_{1,x^+} +\nabla_{\bot}^2 t_{1,x^1} \int dx^- t_{2,x^1}+\nabla_{\bot}^2 t_{1,x^2} \int dx^- t_{2,x^2}\right)\label{T++} \\&
(T_{--})^{(2)}= \frac{1}{4}\frac{ 1}{\kappa_4^2} \int  dx^+\left(  t_{1} \nabla_{\bot}^2 t_{2,x^-}  +\nabla_{\bot}^2 t_{2,x^2} \int dx^+ t_{1,x^2}+\nabla_{\bot}^2 t_{2,x^1} \int dx^+ t_{1,x^1}\right)\label{T--}\\&
    (T_{+1})^{(2)}=              \frac{1}{4}\frac{1}{\kappa_4^2} \nabla_{\bot}^2 t_{1} \int  dx^- t_{2,x^1}
\hspace{0.45in}
   (T_{-1})^{(2)}=              \frac{1}{4}\frac{ 1}{\kappa_4^2} \nabla_{\bot}^2 t_{2} \int  dx^+ t_{1,x^1}         \label{T+-1}\\ &
    (T_{+2})^{(2)} = 	        \frac{1}{4}\frac{ 1}{\kappa_4^2} \nabla_{\bot}^2 t_{1} \int  dx^- t_{2,x^2}
\hspace{0.45in}
 (T_{-2})^{(2)} = 	        \frac{1}{4}\frac{ 1}{\kappa_4^2} \nabla_{\bot}^2 t_{2} \int  dx^+ t_{1,x^2}\label{T+-2}\\&
 (T_{11})^{(2)}= (T_{22})^{(2)}= (T_{12})^{(2)}=0
\end{align}
\end{subequations}
The first equality in equation (\ref{T+-}) is not completely obvious and so we prove it below by considering

\begin{align}\label{Tpm}
 (T_{+-})^{(2)}&=\left(g_{+\mu}g_{-\nu} T^{\mu \nu}\right)^{(2)} \notag\\&
=g_{+\mu}^{(0)}g_{-\nu}^{(1)} (T^{\mu \nu})^{(1)}+g_{+\mu}^{(1)}g_{-\nu}^{(0)} (T^{\mu \nu})^{(1)}+g_{+\mu}^{(0)}g_{-\nu}^{(0)} (T^{\mu \nu})^{(2)}\notag\\&
=g_{+-}^{(0)}g_{--}^{(1)} (T_{++})^{(1)}+g_{++}^{(1)}g_{-+}^{(0)} (T^{--})^{(1)}+g_{+-}^{(0)}g_{-+}^{(0)} (T^{-+})^{(2)}\notag\\&
=(-1)(t_2) \left(-\frac{1}{2 \kappa_4^2} \nabla_{\bot}^2t_1 \right)+(t_2)(-1)\left(-\frac{1}{2 \kappa_4^2} \nabla_{\bot}^2 t_1 \right)\notag\\&
+(-1)(-1) \left( -\frac{1}{4}\frac{ 1}{ \kappa_4^2}     \left(   t_{2} \nabla_{\bot}^2 t_{1} + t_{2} \nabla_{\bot}^2 t_{1} \right) \right)  \notag\\&
= \frac{1}{4}\frac{ 1}{ \kappa_4^2}     \left(   t_{2} \nabla_{\bot}^2 t_{1} + t_{2} \nabla_{\bot}^2 t_{1} \right)
\end{align}
where in the fourth equality we used the fact that the Stress-Energy tensor of the point particles (see (\ref{T1}), (\ref{T2}) and (\ref{nab1})) to first order in $\mu$ may take the form 
\begin{align}\label{T12nab}
T_{++}^{(1)}=-\frac{1}{2\kappa_4^2} \nabla_{\bot}^2 t_1 \hspace{0.3in} T_{--}^{(1)}=-\frac{1}{2\kappa_4^2} \nabla_{\bot}^2 t_2.
\end{align}
Despite working mostly with (\ref{Tmn2}) which is a compact expression, for concreteness, we write $T_{\mu \nu}$ of (\ref{Tmn2}) in terms of the coordinates in order to clarify its form. Defining 
\begin{align}\label{r12}
 \vec{r_1}=   \vec{r}- \vec{b_1}               \hspace{0.3in}            \vec{r_2}=   \vec{r}- \vec{b_2} .
\end{align}
and employing (\ref{nab1}) in  (\ref{Tmn2}) we obtain

\begin{subequations}\label{Tmnc}
 \begin{align}
&  (T_{+-})^{(2)}=    \frac{\pi \mu^2}{2 \kappa_4^2}\log(2k|b|) \delta(x^+)\delta(x^-)\left( \delta^{(2)}(\vec{r_1})  +   \delta^{(2)}(\vec{r_2})) \right) \label{T+-c}   \\&
(T_{++})^{(2)}=  \frac{\pi \mu^2}{2 \kappa_4^2} \theta(x^-)\Bigg[\log(2k|b|) \delta'(x^+)\delta^{(2)}(\vec{r_1}) +\frac{x^-}{x^1+b}\delta(x^+)\delta'(x^1-b)\delta(x^2) \notag\\&
\hspace{1.55in}+\frac{x^- x^2}{4b^2+(x^2)^2}\delta(x^+)\delta(x^1-b)\delta'(x^2) \Bigg] \label{T++c}\\&
    (T_{+1})^{(2)}= \frac{\pi \mu^2}{4 \kappa_4^2 |b|} \theta(x^-)\delta(x^+)\delta^{(2)}(\vec{r_1})   \label{T+-1c}\\ &
    (T_{+2})^{(2)} = 	     0 \label{T+2c}
%
\end{align}
\end{subequations}
The asymmetry between $T_{+1}^{(2)}$ and $T_{+2}^{(2)}$ is due to the fact that the impact parameter $\vec{b}$ has only $x^1$ component (see (\ref{s12})). The rest non zero components that complete (\ref{Tmnc}) may be obtained using the discrete symmetries of the problem: $T_{--}^{(2)}$ and $T_{-1}^{(2)}$ may be obtained from $T_{++}^{(2)}$ and $T_{+1}^{(2)}$ respectively by interchanging $+ \leftrightarrow-$ and $b \leftrightarrow-b$.

We want to justify our claim that the $T_{\mu \nu}^{(2)}$ corrections to $T_{\mu \nu}$ correspond to null geodesics. For this we consider the line element $ds^2=-2dx^+dx^- + t_2 (dx^-)^2+dx_{\bot}$ of the second shockwave $t_2$ (the analogue of (\ref{ds1})). For time-like distances and for fixed transverse position we have $ds^2=0=-2dx^+dx^- + t_2 (dx^-)^2$ and integrating over the discontinuity due to the trajectory of the first particle ($t_1$) along the second shockwave $(t_2)$ we deduce that 

\begin{align}\label{d+}
(\Delta x^+)^{(1)}=\int dx^-=\frac{1}{2} \int dx^- t_2
\end{align}
The superscript $^{(1)}$ on this equation highlights the fact that the discontinuity along the shockwave $t_1$ is of first order in $\mu$ and it implies that the trajectory of the second particle will be modified by along $x^+$ by $(\Delta x^+)^{(1)}$. But according to the argument of the first $\delta$-function of (\ref{TI1}), equation (\ref{d+}) is exactly equal to the shift along $x^+$ that we have already encountered from the geodesic analysis (for the first particle). This completes our argument.

\section{Conservation, Tracelessness and Field Equations}

The second order corrections to (the total) $T_{\mu \nu}$ have already been calculated in the previous section. One, may check by a direct computation using (\ref{nab^2}) that this $T_{\mu \nu}$ is covariantly conserved. Explicitly this means that
\begin{align}\label{con}
& \left((\nabla^{\mu})^{(0)}+ (\nabla^{\mu})^{(1)}\right)   \left( (T_{\mu\nu})^{(1)}+(T_{\mu\nu})^{(2)} \right) =\delta_{\pm \nu} \nabla_{\bot}^2t_1\nabla_{\bot}^2t_2+O(\mu^3)\notag\\&
 \hspace{0.7in} \text{$=O(\mu^3)$ when $\vec{b}_1 \neq \vec{b}_2$ otherwise $\sim \delta_{\pm \nu}\delta^{(2)}(0)$}
\end{align}
where $\nabla$ denotes a covariant derivative while the superscripts denote the order in $\mu$ while we have used the identity (\ref{nab^2}). Therefore we conclude that $T_{\mu \nu}$ is conserved if and only if the impact parameter is not zero \footnote{The zero impact parameter $b$ causes problem in the metric as well. As we will see, as $b\rightarrow 0$, the metric tensor diverges logarithmically (see section \ref{scon}).}. This is one of our main conclusions in this paper.

It is also useful to compute the trace of $T_{\mu \nu}$ as it enters the field equations (see (\ref{Ein})). A short computation yields to
\begin{align}\label{Tr}
T=g^{\mu \nu}T_{\mu \nu}=(g^{\mu \nu})^{(1)}(T_{\mu \nu})^{(1)}+(g^{\mu \nu})^{(0)}(T_{\mu \nu})^{(2)}=0+O(\mu^3)
\end{align}
which shows that the stress-Energy tensor is traceless to order $\mu^2$. Tracelessness is very convenient as it simplifies Einstein's equations which become
\begin{align}\label{EE}
R_{\mu \nu}  = \kappa_4^2
T_{\mu \nu}+O(\mu^3)  \hspace{0.35in}\kappa_4^2=8\pi G_4.
\end{align}

\chapter{Field Equations} \label{FE} 

\section{ Field Equations to $O(\mu^2)$ }\label{FE1}
In this section we wish to write an explicit form of (\ref{EE}) up to order $O(\mu^2)$. In order to determine these (differential) equations we take into account that the zeroth order terms satisfy (\ref{EE}) trivially as $R_{\mu \nu}^{(0)}=T_{\mu \nu}^{(0)}=0$ while $R_{\mu \nu}^{(1)}$ (resulting from the first order terms of (\ref{s12})) is compensated by $T_{\mu \nu}^{(1)}$ of equations (\ref{T1}) and (\ref{T2}). Thus, we only need 
\begin{align}\label{Rmn2}
R_{\mu \nu}^{(2)} = \kappa_4^2T_{\mu \nu}^{(2)}
\end{align}
where $T_{\mu \nu}^{(2)}$ has already been calculated in the previous Chapter and is given by (\ref{Tmn2}). It is crucial to state that $R_{\mu \nu}^{(2)}$ receives two different type of contributions: (a) The contribution due to the (pre)existing shockwaves (that is due to the $t_1$ and $t_2$ terms of (\ref{s12})); we denote this contribution by $(R_{\mu \nu}^{(2)})_{t_{12}}$. (b) The contribution due to the (second order) corrections (in $\mu$) of the metric (that is due to $g_{\mu \nu}^{(2)})$; we denote this contribution by $(R_{\mu \nu}^{(2)})_g$. Recalling equation (\ref{s12}) that gives the form of the metric at all times and expanding (\ref{Rmn2}) to $O(\mu^2)$, we expect that it should have the form 
\begin{align}\label{Rmn2gt}
(R_{\mu \nu}^{(2)})_g+(R_{\mu \nu}^{(2)})_{t_{12}} = \kappa_4^2T_{\mu \nu}^{(2)}
\end{align}
where $(R_{\mu \nu}^{(2)})_{t_{12}}$ (and $T_{\mu \nu}^{(2)}$) is known while $(R_{\mu \nu}^{(2)})_g$ is what we will use in order to determine $g_{\mu \nu}^{(2)})$ (see (\ref{deq1})). Equation (\ref{Rmn2gt}) may also be cast in the form
\begin{align}\label{Rmn2gtT}
(R_{\mu \nu}^{(2)})_g = \kappa_4^2T_{\mu \nu}^{(2)}-(R_{\mu \nu}^{(2)})_{t_{12}}
\end{align}
and view $(R_{\mu \nu}^{(2)})_{t_{12}}$ as an effective (contribution to the tottal) Stress-Energy tensor (see (\ref{deq1}) and (\ref{deq})). Dropping the superscripts $^{(2)}$ from $g_{\mu \nu}^{(2)}$ and $T_{\mu \nu}^{(2)}$  and the superscripts $^{(1)}$ from $t_{1,2}^{(1)}$ for simplicity \footnote {We restore the superscripts that denote the order where is necessary.}, we find that the components of (\ref{EE}) to second order in $\mu$ read
\begin{subequations}\label{deq1}
\begin{align}
(++)\hspace{0.15in} \frac{1}{2}\Big[ &-g_{++,x^1x^1}-g_{++,x^2x^2}+2g_{+1,x^+x^1}+2g_{+2,x^+x^2}
\notag\\&
-g_{11,x^+x^-}-g_{22,x^+x^-}\Big]=\kappa_4^2T_{++},\label{++1}\\
(+-)\hspace{0.15in} \frac{1}{4}\Big[ &-2g_{+-,x^2x^2}-2g_{+-,x^1x^1}+2g_{+2,x^-x^2}+2g_{+1,x^-x^1}-2g_{++,x^-x^-}
\notag\\&
+2g_{-2,x^+x^2}+2g_{-1,x^+x^1}-2g_{11,x^+x^-}-2g_{22,x^+x^-}+4g_{+-,x^+x^-}\notag\\&
-2g_{--,x^+x^+}-2t_{1,x^1}t_{2,x^1}-2t_{1,x^2}t_{2,x^2}+t_{1,x^+}t_{2,x^-}\Big]=\kappa_4^2T_{+-},\label{+-1}\\
(+1)\hspace{0.15in} \frac{1}{4}\Big[ &-2g_{+1,x^2x^2}+2g_{+2,x^+x^-}-2g_{++,x^-x^1}+2g_{12,x^+x^2}-2g_{22,x^+x^1}
\notag\\&
+2g_{+-,x^+x^1}+2g_{+1,x^+x^-}-2g_{-1,x^+x^+}+t_{1,x^+}t_{2,x^1}\Big]=\kappa_4^2T_{+1},\label{+11}\\
(11)\hspace{0.15in} \frac{1}{2}\Big[ &-g_{11,x^2x^2}+2g_{12,x^1x^2}-g_{22,x^1x^1}+2g_{+-,x^1x^1}-2g_{+1,x^-x^1}-2g_{-1,x^+x^1}\notag\\&
+2g_{11,x^+x^-}+t_{1,x^1}t_{2,x^1}+t_{1,x^1x^1}t_{2}+t_{1}t_{2,x^1x^1}\Big]=k_{4}^{2} T_{11}\label{111}=0,\\
(22)\hspace{0.15in} \frac{1}{2}\Big[ &-g_{11,x^2x^2}+2g_{12,x^1x^2}-g_{22,x^1x^1}+2g_{22,x^+x^-}-2g_{+2,x^-x^2}-2g_{-2,x^+x^1}\notag\\&
+2g_{+-,x^2x^2}+t_{1,x^2}t_{2,x^2}+t_{1,x^2x^2}t_{2}+t_{1}t_{2,x^2x^2}\Big]
=k_{4}^{2} T_{22}\label{221}=0,\\
(12)\hspace{0.15in} \frac{1}{4}\Big[ &
4g_{+-,x^1x^2}-2g_{+1,x^-x^2}-2g_{+2,x^-x^1} -2g_{-1,x^+x^2}-2g_{-2,x^+x^1}+4g_{12,x^+x^-}\notag\\&
 t_{2,x^2} t_{1,x^1}+ t_{1,x^2}t_{2,x^1}+2t_{2}t_{1,x^1x^2} +2t_{1}t_{2,x^1x^2} \Big]=\kappa_4^2T_{12}=0\label{121}
\end{align}
\end{subequations}
where $t_{1,2}$ are given by (\ref{s12}) and correspond to the geometry for negative times while the components of $T_{\mu \nu}$ are given by (\ref{Tmn2}). Indeed (\ref{deq1}) has the expected form of equation (\ref{Rmn2gt}). The above set of the field equations has been written without specifying the gauge. In the next section, we will see how these may be simplified by making a convenient gauge choice.

\section{Choosing the Gauge}\label{CG}

We follow the standard procedure in order to solve (\ref{deq1}): We define a new coordinate system $\tilde{x}^{\nu}$ with respect to the old one $x^{\nu}$ (see (\ref{LC})) by

\begin{align}\label{ncs}
\tilde{x}^{\nu}=x^{\nu}+(\xi^{\nu})^{(2)}
\end{align}
where $(\xi^{\nu})^{(2)}$ is an arbitrary function of the (old coordinates) $x^{\kappa}$ and is second order in $\mu$. Obviously, this transformation induces a second order change in $\mu$ to $g_{\mu \nu}^{(2)}$ but does not alter $g_{\mu \nu}^{(0)}$ and $g_{\mu \nu}^{(1)}$. More precisely the second order terms of the metric transform to

\begin{align}\label{Nm}
\tilde{g}_{\mu \nu}^{(2)}=g_{\mu \nu}^{(2)}+\xi_{\mu;\nu}^{(2)}+\xi_{\nu;\mu}^{(2)} 
\end{align}
where the semicolon denotes a covariant derivative\footnote{In fact $\tilde{g}_{\mu \nu}^{(2)}-g_{\mu \nu}^{(2)}=\xi_{\mu;\nu}+\xi_{\nu;\mu}$ is exactly equal to the action of the Lie derivative acting on $g_{\mu \nu}^{(2)}$ along the vector field $\xi^{\nu}$, i.e. $\cal{L}_{\xi}$$g_{\mu \nu}^{(2)}=\xi_{\mu;\nu}^{(2)}+\xi_{\nu;\mu}^{(2)}$.}. What is remarkable is that the field equations remain invariant under this transformation as it may be shown on general grounds \cite{DEath:1992hb,Papa:1974,Carroll:1997ar,Weinberg:1972}. For concreteness, we exhibit it here for the $(12)$ component, equation (\ref{121}). Taking into account that at the order we are working, the covariant derivative may be replaced by ordinary differentiation, plugging the tensor $\xi_{\mu;\nu}^{(2)}+\xi_{\nu;\mu}^{(2)} $ into the differential part of (\ref{121}) and dropping (again) the superscript $^{(2)}$ from the $\xi$'s for simplicity, we obtain
\begin{align}\label{xis}
  \frac{1}{4}& \Big[ 
4\left(\xi_{+,x^-}+\xi_{-,x^+} \right)_{,x^1x^2}-2\left(\xi_{+,x^1}+\xi_{1,x^+} \right)_{,x^-x^2}-2\left(\xi_{+,x^2}+\xi_{2,x^+} \right)_{,x^-x^1} \notag\\&
-2\left(\xi_{-,x^1}+\xi_{1,x^-} \right)_{,x^+x^2}-2\left(\xi_{-,x^2}+\xi_{2,x^-} \right)_{,x^+x^1}+4\left(\xi_{1,x^2}+\xi_{2,x^1} \right)_{,x^+x^-}\Big] \notag\\&
 =\frac{1}{4}\Big[\left\{  \left(4\xi_{+,x^-,x^1,x^2}-2\xi_{+,x^1,x^-,x^2}-2\xi_{+,x^2,x^-,x^1}\right) +(+\leftrightarrow-) \right \}\notag\\&
\hspace{0.25in}+\left\{ \left(4\xi_{1,x^2,x^+,x^-}-2\xi_{1,x^+,x^-,x^2}-2\xi_{1,x^-,x^+,x^2}\right)+(1\leftrightarrow2)\right\}
\Big] =0
\end{align}
where we used the fact that partial derivatives commute. 
So far, the vector field $\xi$ has been arbitrary while the result of the transformation (\ref{ncs}) on (\ref{deq1}) is just the relabeling $ g_{\mu \nu}\rightarrow \tilde{g}_{\mu \nu}$. A convenient choice of $\xi_{\nu}$ is the one that  satisfies de Donder gauge
\begin{align}\label{gc}
\tilde{g}_{\mu \nu},^{\mu}-\frac{1}{2}\eta_{\mu \nu}\tilde{g}_{\kappa}\hspace{0.01in}^{\kappa}=0
\end{align}
where $\eta_{\mu \nu}=g_{\mu \nu}^{(0)}$ is the flat metric. Applying this gauge to the field equations that we are interested, equations (\ref{deq1}), and dropping the tilde symbol from $\tilde{g}_{\mu \nu}$ for simplicity, the field equations simplify to
\begin{subequations}\label{deq}
\begin{align}
(++)\hspace{0.4in}& \Box g_{++}=- \frac{1}{2}  \int  dx^-\Bigg(  t_{2}  \nabla_{\bot}^2 t_{1,x^+} \notag\\&
\hspace{1.2in}   +\nabla_{\bot}^2 t_{1,x^1} \int dx^- t_{2,x^1}+\nabla_{\bot}^2 t_{1,x^2} \int dx^- t_{2,x^2}\Bigg),\label{++}\\
(+-)\hspace{0.4in}&  \Box g_{\mu \nu}=- \frac{1}{2}\left(   t_{2} \nabla_{\bot}^2 t_{1} + t_{2} \nabla_{\bot}^2 t_{1} \right)-t_{1,x^1}t_{2,x^1}\notag\\&
\hspace{1.4in}-t_{1,x^2}t_{2,x^2}+\frac{1}{2}t_{1,x^+}t_{2,x^-},\label{+-}\\
(+1)\hspace{0.46in}& \Box g_{++}=-\frac{1}{2} \nabla_{\bot}^2 t_{1} \int  dx^- t_{2,x^1}+\frac{1}{2}t_{1,x^+}t_{2,x^1},\label{+1}\\
(11)\hspace{0.5in}& \Box g_{11}=t_{1,x^1}t_{2,x^1} + t_{1,x^1x^1}t_{2} + t_{1}t_{2,x^1x^1}=0,\label{11}\\
%
%
%
(12)\hspace{0.5in}  & \Box g_{12}=
\frac{1}{2} t_{2,x^2} t_{1,x^1}+ \frac{1}{2}  t_{1,x^2}t_{2,x^1}+t_{2}t_{1,x^1x^2} +t_{1}t_{2,x^1x^2} =0\label{12}
\end{align}
\end{subequations}
where we have used (\ref{Tmn2}) while $\Box$ is the scalar operator in flat space; that is
\begin{align}\label{box}
\Box \equiv \eta^{\mu \nu}\partial_{\mu} \partial _{\nu}=-2 \partial_{x^+}\partial_{x^-}+\nabla_{\bot}^2.
\end{align}
In the next chapter we will see how equations (\ref{deq}) may be solved imposing appropriate boundary conditions.

\chapter{Solving the Field Equations and Causality} \label{SEq} 

\section{Green's Function and Boundary Conditions}\label{seq1}
In this section we will show how to solve (\ref{deq}) by seeking for causal solutions. The casual boundary conditions imply that the flat metric in the presence of the shockwaves, as can be checked, is an exact solution to Einsteins equations with a right hand side given by (\ref{T1}) and (\ref{T2}), for negative times only. For positive times, that is for $x^+>0$ and $x^->0$, the second order corrections (in $\mu$) of the metric are switched on as the point $x^+=x^-=0$ is the collision point. Simultaneously, the initial Stress-Energy tensor of the (massless) particles that induces the shock-waves suffers a change (see figure \ref{SelfInt4d} and (\ref{Tmn2})) that also has to be taken into account. 

The retarded Green's function we need corresponding to the differential operator (\ref{box}) is known and in light-cone coordinates is given by

\begin{align}\label{GF}
G(x^{\mu} - x'^{\mu})=-\frac{1}{4\pi}\frac{\theta(x^+-x'^+)\theta(x^--x'^-)}{\frac{1}{\sqrt{2}} \left((x^+-x'^+)+(x^--x'^-)\right)}
\delta \left(\sqrt{2(x^+-x'^+)(x^--x'^-)}-|\vec{r}-\vec{r'}| \right)
\end{align}
where according to (\ref{s1}), $\vec{r}=(x^1,x^2)$, $\theta$ denotes a theta (step) function while equation (\ref{GF}) (the retarded Green's function) satisfies

\begin{align}\label{boxGF}
\Box G(x^{\mu} - x'^{\mu})=\delta(x^+-x'^+)\delta(x^--x'^-)\delta^{(2)}(\vec{r}-\vec{r'}).
\end{align}

\section{Integration over the Light-Cone Plane}\label{LCP}

The procedure we have to follow is standard: we convolute the right hand sides of (\ref{deq}) with (\ref{GF}) and integrate in all over space-time. 
We find it convenient to introduce the following notation

\begin{align}\label{sh}
 G\otimes f(x'^{\kappa})\equiv \int_{-\infty}^{\infty}dx'^+ \int_{-\infty}^{\infty}dx'^- \int d^2\vec{r'} G(x^{\mu} - x'^{\mu}) f(x'^{\kappa})
\end{align}
where $f(x^{\kappa})$ is any arbitrary function of $x^{\kappa}$ while the last integral denotes integration in the transverse plane. We wish to perform the $x^{\pm}$ integrations for all the possible cases that we will encounter while specifying $g_{\mu \nu}^{(2)}$ from (\ref{deq}). We organize these integrations in five cases while we leave the details of the calculation for Appendix \ref{A}.

\vspace{0.1in}
\underline{Remark 1:}
\vspace{0.1in}
In addition to the five cases of the $x^{\pm}$ integrations which we solve in Appendix \ref{A} there exists another case that involves the evaluation of
\begin{align}\label{c6}
G\otimes \int dx^-\int dx^- \Big(t_{2,x^1} \nabla_{\bot}^2 t_{1,x^1}  +  t_{2,x^2} \nabla_{\bot}^2 t_{1,x^2} \Big) 
\end{align}
that arises from (\ref{++}). Due to the complication of the calculation we evaluate this term in a separate Appendix ( see Appendix \ref{B}, equation (\ref{A2})).

\vspace{0.1in}
\underline{Remark 2:}
\vspace{0.1in}
The results of all of the integrations in the light-cone plane (performed in Appendices \ref{A} and \ref{B}) are proportional to the product $\theta(x^+)\theta(x^-)$. This implies that the second order corrections to $g_{\mu \nu}$ appear in the forward light-cone (see figure \ref{etatau}) which is what we have initially demanded by seeking for a causal solution.

\vspace{0.1in}
\underline{Remark 3:}
\vspace{0.1in}
The right hand sides of (\ref{deq}) contains expressions of the form $t_1t_2$ differentiated with respect to $x^{\mu}$ in some fashion. According to (\ref{s12}), these expressions are proportional to $\delta(x^+)\delta(x^-)$ or their derivatives. Our previous analysis has already taken care of the $x^{\pm}$ integrations and so from now on, by $t_{1,2}$ we will mean just the transverse part of $t_{1,2}$: $-\mu \log(\sqrt{(x^1\pm b)^2+(x^2)^2})$.

\section{Integration over the Transverse Plane}\label{TP}

Having performed the $x^{\pm}$ integrations we move to the integration over the transverse plane. The quantities we have to integrate have the structure \footnote{There is another case where we have to integrate terms of the form $\nabla_{\bot}^2 t_{1,2}$. However, according to (\ref{nab1}) these (transverse) integrations are trivial as they involve delta functions.} $\partial_{x_a^i}(t_1t_2)$ or $\partial^2_{x_a^i x_c^j}(t_1t_2)$ where $a,c;i,j=1,2$. The subscript $a$ (c) and the superscript $i$ $(j)$ denotes differentiation of the source $t_a$ $(t_c)$ with respect to the space-time coordinate $x^i$ $(x^j)$. We may reduce the number of the different integrals that we have to perform by working as follows. We firstly introduce the vectors
\begin{align}\label{b}
\vec{b_1}=(b_{11},b_{12}) \hspace{0.3in}\vec{b_2}=(b_{21},b_{22}) 
\end{align}
and generalize the form of (the transverse part of) $t_{1,2}$ given by (\ref{s12}) to
\begin{align}\label{nt}
t_1(\vec{r}-\vec{b_1})=-\mu \log(k r_1)  \hspace{0.3in}  t_2(\vec{r}-\vec{b_2})=-\mu \log(k r_2)
\end{align}
where $(\vec{r}_{1,2}$ where defined by (\ref{r12}). The next step is to exchange the derivatives acting on $t_{1,2}$, that is $\partial_{x_a^i}$ with differentiations with respect to $b$'s of (\ref{b}), that is with $-\partial_{b_{a i}}$\footnote{So for instance $t_{1,x^1}t_{2,x^1}$ takes the form $\partial^2_{b_{11} b_{21}}(t_1t_2)$.}. Finally, at the end of our calculations we take the limits
\begin{align}\label{lim}
 \vec{b_1} \rightarrow  (b,0) \hspace{0.3in}  \vec{b_2}\rightarrow  (-b,0). 
\end{align}
Looking equations (\ref{+-})-(\ref{12}) we see that they involve the product $t_1t_2$ differentiated with respect to the transverse coordinates \footnote{Where $t_1t_2 \sim \log(k r_1) \log(k r_2)$; see (\ref{s12}) while the $x^{\pm}$ contributions have already been taken into account in the previous section.}. 
Exchanging the transverse differentiations, according to our earlier discussion in this section, with derivatives with respect to the components of $\vec{b}_{1,2}$  and taking into account the transverse part of the Green's function, (\ref{GF}), we see at once that we have to calculate the following integral
\begin{align}\label{Jin}
{\cal J} (r_1,r_2,\tau)=\frac{1}{2 \pi \tau}\int_{0}^{\infty}\int_{0}^{2\pi}r'dr'd\phi' \delta(\tau-r') \log(k |\vec{r'}+\vec{r_1}|) \log(k |\vec{r'}+\vec{r_2}|)
\end{align}
where we have introduced the convenient factor $\frac{1}{2\pi \tau}$.
Now, the non trivial integration is the angular integration as the radial one becomes trivial due to the $\delta$-function. Both of the integrations are performed in Appendix \ref{C} and the final result reads
\begin{align} \label{J}
{\cal J} (r_1,r_2,\tau)& = \theta(r_1-\tau)\theta(r_2-\tau){\cal J}_1 (r_1,r_2,\tau)+ \theta(\tau-r_2)\theta(r_1-\tau){\cal J}_2 (r_1,r_2,\tau) \notag\\&
+ \theta(\tau-r_1)\theta(r_2-\tau){\cal J}_3 (r_1,r_2,\tau)+\theta(\tau-r_1)\theta(\tau-r_1){\cal J}_4 (r_1,r_2,\tau)
\end{align}
where the ${\cal J}$'s may be found with the help of table \ref{ta1} and equation (\ref{Ja}). Equation (\ref{J}) is the last ingredient that allows us to obtain the desired solutions for equations (\ref{deq}). We display the results in the next section.

\section{The Formula for $g_{\mu \nu}^{(2)}$}\label{FGmn}
%
Having performed all of the integrations arising from the convolution of the right hand sides of (\ref{deq}) with the Green's function (\ref{GF}), we are in a position to derive the final formulas for $g_{\mu \nu}^{(2)}$ 
\footnote{We have restored the superscript $^{(2)}$ on the corrections of $g_{\mu \nu}$ in order to highlight the order in $\mu$ that we are working.}. The ingredients that we need, have been obtained or defined in the previous sections and in Appendices \ref{A}, \ref{B} and \ref{C}. To begin with, we need the defining equations for $t_{1,2}$ and $T_{\mu \nu}^{(2)}$ given by (\ref{s12}) and (\ref{Tmn2}) respectively but with the generalized $\vec{b}_{1,2}$ (see (\ref{b})) instead (of (\ref{lim})) while identity (\ref{nab1}) is very useful.
 We also need the value of the integral $\cal{J}$ defined in (\ref{Jin}) and given by (\ref{J}), (\ref{Ja}) and table \ref{ta1} as well as (\ref{r12}) and (\ref{tao}) that define $r_{1,2}$ and $\tau$, $\eta$ respectively. The final formula for  $g_{\mu \nu}^{(2)}$ is eventually given below.
 
\vspace{0.3in}
\underline{The Formula for $g_{\mu \nu}^{(2)}$}
\vspace{0.1in}
\begin{subequations}\label{gmn2}
\begin{align}
 g_{++}^{(2)} &=\lim_{\vec{b}_{1,2}\to (\pm b,0)}\Bigg \{ \frac{1}{\sqrt{2}} \mu^2 \theta(x^+)\theta(x^-) \Bigg \{                   
 \log \left(k|\vec{b}_2-\vec{b}_1|\right) \partial_{x^+} \left(  \frac{r_1}{r_1^2+2 (x^{\pm})^2} \theta(\tau-r_1) \right) \notag\\&
+\frac{1}{2 x^+}\left[ \frac{b_{11}-b_{21}}{|\vec{b}_2-\vec{b}_1|^2}\theta(\tau-r_1)  \partial_{x^1}  \left(r_1 \frac{\tau^2-r_1^2}{r_1^2+2 (x^+)^2} \right)+ \big(1 \leftrightarrow 2 \big) \right]   \Bigg\}\Bigg\}  ,                         \label{g++2} \\
     g_{+-}^{(2)} & \lim_{\vec{b}_{1,2}\to (\pm b,0)}\Bigg \{ 
     \frac{1}{2} \mu^2 \theta(x^+)\theta(x^-) \text{sech} \hspace{0.02in} \eta  \Bigg\{\frac{1}{2 \tau}  \log \left(k|\vec{b}_2-\vec{b}_1|\right)\delta(\tau-r_1) \notag\\&
+\left[    \partial^2_{b_{11}b_{21}}-\frac{1}{4} \left(\frac{1}{\tau^2}\text{sech$^2$}\hspace{0.02in}\eta+\frac{1}{2}\tau \partial_{\tau}  \left(\frac{1}{\tau}\partial_{\tau}           \right) \right)    \right]{\cal J} (r_1,r_2,\tau) 
+\Big( 1\leftrightarrow2 \Big) \Bigg\} \Bigg\} , \label{+-2} \\
  g_{+1}^{(2)} &=\lim_{\vec{b}_{1,2}\to (\pm b,0)}\Bigg \{ 
  \frac{1}{\sqrt{2}} \mu^2 \theta(x^+)\theta(x^-) \Bigg \{    \frac{b_{11}-b_{21}}{|\vec{b}_2-\vec{b}_1|^2}\frac{r_1}{r_1^2+2 (x^{\pm})^2} \theta(\tau-r_1)    \notag\\&
 \hspace{1.20in}+\frac{1}{2}  (\partial_{b_{21}})   \left [\frac{1}{1+e^{ \pm 2\eta}}\partial_{\tau} -\frac{1}{2 \tau} \text{sech}^2\hspace {0.02in}\eta  \right ] {\cal J} (r_1,r_2,\tau)    \Bigg\}  \Bigg\} , \label{g+12} \\
 g_{11}^{(2)} &=\lim_{\vec{b}_{1,2}\to (\pm b,0)}\Bigg \{ 
 -\frac{1}{2} \mu^2 \theta(x^+)\theta(x^-) \text{sech} \hspace {0.02in}\eta \notag\\&
 \hspace{1.8in}\times \Big\{ \partial_{b_{11}b_{21}}^2 +\partial_{b_{11}b_{11}}^2 +  \partial_{b_{21}b_{21}}^2  \Big\}   {\cal J} (r_1,r_2,\tau)  \Bigg\}          ,                      \label{g112} \\
 g_{12}^{(2)} &=\lim_{\vec{b}_{1,2}\to (\pm b,0)}\Bigg \{ 
 -\frac{1}{4} \mu^2 \theta(x^+)\theta(x^-) \text{sech}\hspace {0.02in}\eta\notag\\&
 \hspace{1.2in}  \Big\{ \partial_{b_{22}b_{11}}^2 +\partial_{b_{12}b_{21}}^2 +2\partial_{b_{11}b_{12}}^2  +2\partial_{b_{21}b_{22}}^2  \Big\}   {\cal J} (r_1,r_2,\tau)          \Bigg\}   .                             \label{g122} 
\end{align}
\end{subequations}

In order to arrive to (\ref{gmn2}) we have convoluted (\ref{GF}) with the right hand side of (\ref{deq}) and employed (\ref{c1}), (\ref{c2}), (\ref{c3}), (\ref{c4}), (\ref{c5}) and (\ref{A2}). In particular we have applied (\ref{c5}) and (\ref{A2}) for (\ref{++}), (\ref{c1}) and (\ref{c2}) for (\ref{+-}), (\ref{c3}) and (\ref{c4}) for (\ref{+1}) and (\ref{c1}) for both (\ref{11}) and (\ref{12}).

The reason in preferring to work with the generalized $\vec{b}_{1,2}$ is because the  $g_{\mu 2}^{(2)}$ may be obtain from  $g_{\mu 1}^{(2)}$ under $1 \leftrightarrow 2$ before taking the limits as in (\ref{lim}); thus reducing the amount of calculations.
Finally, $(-\mu)$ are obtained from $(+\mu)$ under the (simultaneous) interchanges $(+ \leftrightarrow -)$ and $(\vec{b}_1 \leftrightarrow \vec{b}_2)$. These steps complete the determination of $g_{\mu \nu}^{(2)}$. Formula (\ref{gmn2}) is the final result of this project and we will analyze it in the next Chapter.
\chapter{Area of Validity and  Conclusions}\label{Conc}

\section {Area of Validity}\label{av}
\begin{figure}
\centering
\includegraphics[scale=0.7]{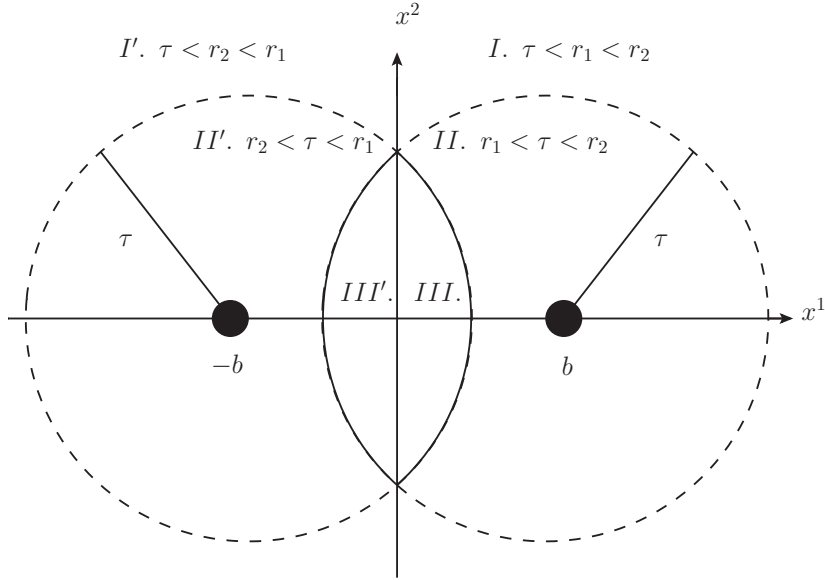}
\caption{The reaction plane: Regions $III$ and $III$$'$ correspond to $r_1 < r_2 < \tau$ and $r_2 < r_1 < \tau$ respectively. The dark dots are the centers of the shockwaves and are located at an impact parameter 2b apart while ${r}_{1,2}$ denote the distance of the arbitrary point $\vec{r}$ from the center of each shockwave (right and left respectively). According to causality, at any given proper time $\tau$, the propagation from the centers will reach the points on the peripheries (at most). This suggests that any given point $\vec{r}$ on the transverse plane of the ``produced" metric at given $\tau$ will evolve according to the region where it belongs to (see equations (\ref{gmn2}), (\ref{Ja}) and (\ref{Jap})) and there are three different possibilities (and three from their mirror images).}
\label{re}
\end{figure}

We have seen below (\ref{s1}) that the parameter $\mu$ we use in our expansion has dimensions of length. On the other hand we know that the components of the metric should be dimensionless. Hence, each power in $\mu$ is compensated by an inverse power of the coordinates times a logarithm (with argument $k r_{1,2}$, $kb$ or $k\tau$) in some power at most. For simplicity (although not necessary), we restrict our discussion at middle -rapidity where $\eta \approx 0$ that is for $x^+ \approx x^-$ (see (\ref{tao}) and figure \ref{etatau}). This means that any $k^{th}$-order (in $\mu$) contribution to the metric, where $k$ is a positive integer, will generally have the form \footnote{This form applies away from the light-cone because on the light-cone there also exist $\delta(\tau-r_{1,2})$ terms (see (\ref{GMN})).}
\begin{align}
g_{\mu \nu}^{(k)}(x^{\mu})=\mu^k \theta(x^+)\theta(x^-) \sum_{ijl}c_{ijk}(x^{\mu})\frac{1}{r_1^i r_2^j \tau^l} \hspace{0.3in}i+j+l=k>0
\end{align}
(see (\ref{gmn2}))where $c_{ijk}(x^{\mu})$ are dimensionless real functions of $x^{\mu}$ \footnote{Such as $\theta$-functions, logarithms and real coefficients}. We thus believe that our expansion is valid at high energies, that is small $\mu$, compared to $r_{1,2}$ and $\tau$. In fact, only one of $r_{1,2}$ or $\tau$ has to be large compared to $\mu$. So for instance (\ref{gmn2}) is a good approximation for small $\tau$ but large $r_{1,2}$ (see figure \ref{re}, region $I$) and also for small $r_{1,2}$ but large $\tau$ (region $III$) provided that the massless particles creating the shockwaves are not energetically enough ($\mu$ is small). This is in contrast to \cite{Taliotis:2010pi} where $\mu$ has dimensions of length to the negative third power and hence the expansion there was valid for early proper times.

\section{Conclussions}\label{scon}

In this work we have found the first non-trivial causal corrections to the problem of shockwaves collisions in gravity created by boosting two black holes to the speed of light. The collision is assumed asymmetric and occurs at low energies (see previous section). In terms of Feynman diagrams, our result, formula (\ref{gmn2}), corresponds to the resummation of the diagrams of figures \ref{interaction} and \ref{SelfInt4d}. 

\vspace{0.3in}

Our concussions are summarized as follows.

\vspace{0.3in}

1. The corrections to $g_{\mu \nu}$ evolve non trivially and constrained by causality in an intuitive way. In particular, the behavior of $g_{\mu \nu}$ at any point on the transverse plane, is determined from whether the propagation from the center of each individual nucleus has enough proper time to reach the point under consideration or not. Figure \ref{re} is a snapshot taken at given proper time $\tau$ and depicts the six kinematical regions where $g_{\mu \nu}$ evolves differently while it has the general form

\begin{align}\label{GMN}
g_{\mu \nu}^{(2)}&=\theta(r_2-\tau)\theta(r_1-\tau)A_{\mu \nu}^I(x^{\kappa},b) +  \theta(\tau-r_1)\theta(\tau-r_2)A_{\mu \nu}^{III}(x^{\kappa},b) \notag\\&
  + \left\{ \theta(r_2-\tau)\theta(\tau-r_1)A_{\mu \nu}^{II}(x^{\kappa},b)+\big(b \leftrightarrow -b \big) \right\}\notag\\&
  +\text{terms proportional to $\delta(\tau-r_{1,2})$} .
\end{align}
The indices $I$, $II$, $III$ on $A_{\mu \nu}$ correspond to the regions  $I$, $II$, $III$ of figure \ref{re} respectively \footnote{The  $(b \leftrightarrow -b)$ terms cover  region $I$$'$,$II$$'$ and $III$$'$ since under this interchange we have $r_1 \leftrightarrow r_2$.}. The $\delta$-function terms arise from differentiating the $\theta$-functions of the right hand sides of (\ref{gmn2}) (see also (\ref{J})). They represent two shockwaves centered at the center of each shockwave and expanding on the transverse plane with speed $\tau$ (that is with the speed of light). This particular behavior of the metric was our initial motivation for dealing with this problem and our calculations confirm our earlier conjecture \cite{Taliotis:2010pi}.

\vspace{0.3in}

2. The presence of matter, $T_{\mu \nu}$, and the back-reactions affect the metric (see for example (\ref{g++2}) and (\ref{g+12})) not only on the forward light-cone but also inside. This implies that we cannot in principle solve Einstein's equations in vacuum (ignoring the point-like particles that create the shocks) arguing that we are away from the sources unless we know the boundary conditions that these sources enforce on the metric inside the light-cone. This is in analogy to classical electrodynamics: solving Laplace equation for the scalar potential away from a point charge sitting at the origin without specifying the boundary conditions, one may obtain the trivial (zero) solution which obviously is not the correct one.

\vspace{0.3in}

3. The presence of the impact parameter $b$ is a necessary requirement and not an additional complication introduced in the problem. Mathematically this is obvious from the fact that both $T_{\mu \nu}$ and $g_{\mu \nu}$ diverge when the impact parameter $b$ tends zero. From equation \ref{con} we have seen that conservation of $T_{\mu \nu}$ is violated violently and behaves as $\nabla ^{\mu}T_{\mu \nu}\sim \delta_{\pm \nu}\delta(0) \neq 0$. The metric tensor also exhibits a problematic behavior in the zero impact parameter limit: the the formula for $g_{\mu \nu}^{(2)}$, equation (\ref{FGmn}), diverges logarithmically when $b \rightarrow 0$ as it is evident (for instance) from equation (\ref{g++2}). A similar ultraviolet (UV) divergence appears in perturbation theory \cite{Kovchegov:1997ke} of gauge theories. This suggests that a head on collision may not be investigated using classical gravity. Instead, one has to apply a quantum theory of gravity in the same way one can not predict the electron-positron annihilation (in head on collisions) using Maxwell's equations. One has to turn into Quantum Electrodynamics in order to describe the process and predict the production of two photons. One may argue that a black hole may be formed and hence hide the violation of conservation behind the event horizon.

\vspace{0.3in}

4. For future projects we propose that one could plot $g_{\mu \nu}^{(2)}$ for several (fixed) impact parameters as a function of $\tau$, $x^1$ and $x^2$ (at central rapidities where $x^+=x^-$) in order to visualize the evolution of the metric. A very important aspect we ignored in our analysis is the role of the ultraviolet cutoff $k$ (see (\ref{s1})) which (for the case of a single shockwave)  seems to define an ergoregion \footnote{We thank Samir Mathur for a related and informative discussion.} (with radius $r_e=1/k$)  on the transverse plane. It would be interesting to check how our solution gets modified for impact parameters $b<1/k$ which would imply that the two ergoregions of the shockwaves overlap. Finally, one could  compute the gravitational radiation, take the limit of $b \rightarrow 0$ and compare the result with the one obtained by \cite{DEath:1992hb,DEath:1992hd,DEath:1992qu}.

\backmatter
\appendix
\chapter{Integration Over the Light-Cone Plane}\label{A}

In this appendix we perform the $x^{\pm}$ part of the integrations resulting when the Green's function (\ref{GF}) acts on the right hand side of (\ref{deq}). Proceeding as in (\ref{sh}) we find out that we have to deal with five different cases (there also exists a sixth case that is calculated in appendix \ref{B}).

\vspace{0.3in}
\hspace{0.25in}\underline{Case I: $\delta(x^+)\delta(x^-)$ terms}
\vspace{0.3in}

\begin{figure}
\centering
\includegraphics[scale=0.635]{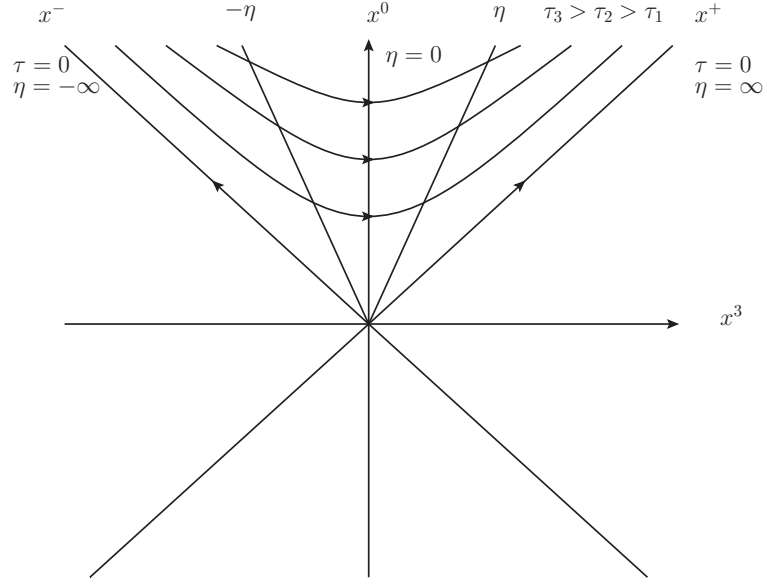}
\caption{The kinematical variables $\tau$ and $\eta$. The hyperbolas indicate curves of constant $\tau$ and increase along $x^0$. The straight lines are lines of constant $\eta$ and increase from left to right as the arrows indicate. Along $x^{\pm}$ we have $(\tau=0, \eta=\pm \infty)$ while along $x^0$ we have $\eta=0$ and $x^+=x^-$.}
\label{etatau}
\end{figure}

This case is trivial and almost all terms of (\ref{deq}) behave in this way. Defining
\begin{align}\label{tao}
\tau=\sqrt{2x^+x^-} \hspace{0.3in}\eta=\frac{1}{2}\log(\frac{x^+}{x^-})\hspace{0.3in}; \hspace{0.3in}x^{\pm}=\frac{1}{\sqrt{2}} \tau e^{\pm \eta}
\end{align}
where $\tau$ is the proper time and $\eta$ the rapidity (see figure \ref{etatau} for a geometrical meaning) we have that
\begin{align}\label{c1}
 G\otimes \delta(x'^+)\delta(x'^-) f_{I}(\vec{r'})=-\frac{1}{4\pi\tau}\theta(x^+)\theta(x^-)
 \text{sech}\hspace{0.02in}\eta \int d^2\vec{r'}  \delta \left(\tau-r' \right)f_{I}(\vec{r}+\vec{r'})
\end{align}
where we have shifted the integration variable setting $\vec{r'}\rightarrow \vec{r'}+\vec{r}$.
%

\vspace{0.2in}
\underline{Case II: $\delta'(x^+)\delta'(x^-)$ terms}
\vspace{0.2in}

This case is more complicated as we have to integrate by parts the $\delta$-functions. Three kind of terms will appear: (a) Terms that differentiate the $\theta(x'^{\pm}-x'^{\pm})$ terms of (\ref{GF}) and hence produce $\delta(x'^{\pm}-x'^{\pm})$ (terms). But the presence $\delta(x'^{\pm}-x'^{\pm})$ forces the $\delta$-function term appearing in (\ref{GF}) to become $\delta(-|\vec{r}-\vec{r'}|)$ which is zero. Hence these terms do not contribute.
(b) We have terms that differentiate the denominator and these contribute to the integrations. (c) Finally, we have terms that either differentiate the $\delta$-function of (\ref{GF}) only or both, the $\delta$-function and the denominator. In order to evaluate these terms we exchange the $\partial_{x'^{\pm}}$ that act on $\delta \left(\sqrt{2(x^+-x'^+)(x^--x'^-)}-|\vec{r}-\vec{r'}| \right)$ with $-\partial_{x^{\pm}}$. Shifting the transverse variable as in the previous case and performing the $x'^{\pm}$ integrations we find that the contribution of both the (b) and (c) terms is%
\begin{align}\label{c2}
 G\otimes \delta'(x'^+)\delta'(x'^-) f_{II}(\vec{r'})&=-\frac{1}{4\pi \tau}\theta(x^+)\theta(x^-) \text{sech}\hspace{0.02in}\eta \notag\\&
\times \left[\frac{1}{\tau^2}\text{sech$^2$}\hspace{0.02in}\eta+\frac{1}{2}\tau \partial_{\tau}  \left(\frac{1}{\tau}\partial_{\tau}           \right) \right]            \int d^2\vec{r'}  \delta \left(\tau-r' \right)f_{II}(\vec{r}+\vec{r'})
\end{align}
where the differential operator $ \frac{1}{\tau^2}\text{sech$^2$}\hspace{0.02in}\eta+\frac{1}{2}\tau \partial_{\tau}  \left(\frac{1}{\tau}\partial_{\tau}           \right)$ acts on the integral while $\partial_{\tau}$ denotes a partial differentiation with respect to $\tau$.

\vspace{0.5in}
\underline{Case III: $\delta'(x^{\pm})\delta(x^{\mp})$ terms}
\vspace{0.3in}

This is a simpler version of case $II$ and working in a similar fashion yields
\begin{align}\label{c3}
 G\otimes \delta'(x'^{\pm})\delta(x'^{\mp}) f_{III}(\vec{r'})&=-\frac{1}{2\pi \tau} \frac{1}{\sqrt{2}}\theta(x^+)\theta(x^-)  \notag\\&
\times \left [\frac{1}{1+e^{ \pm 2\eta}}\partial_{\tau} -\frac{1}{2 \tau} \text{sech}^2\hspace {0.02in}\eta  \right ]            \int d^2\vec{r'}  \delta \left(\tau-r' \right)f_{III}(\vec{r}+\vec{r'}).
\end{align}

\vspace{0.1in}
\underline{Case IV: $\delta(x^{\pm})\theta(x^{\mp}) $ terms}

%
\begin{align}\label{c4}
 G&\otimes \delta(x'^{\pm})\theta(x'^{\mp}) f_{VI}(\vec{r'})=-\frac{1}{4\pi } \theta(x^{\pm})  \notag\\&
\times \int_{-\infty}^{\infty}dx'^{\mp} \theta(x'^{\mp}) \theta(x^{\mp}-x'^{\mp})        \int d^2\vec{r'}      \frac{\delta \left(\sqrt{2 x^{\pm}(x^{\mp}-x'^{\mp})}- r'\right)} {\frac{1}{\sqrt{2}}\left( x^{\pm}+(x^{\mp}-x'^{\mp}) \right)}        f_{IV}(\vec{r}+\vec{r'})\notag\\&
=-\frac{1}{4\pi } \theta(x^+)\theta(x^-) \int_{0}^{x^{\mp}}dx'^{\mp}       \int d^2\vec{r'}      \frac{\delta \left(\sqrt{2 x^{\pm}(x^{\mp}-x'^{\mp})}- r'\right)} {\frac{1}{\sqrt{2}}\left( x^{\pm}+(x^{\mp}-x'^{\mp}) \right)}        f_{VI}(\vec{r}+\vec{r'})\notag\\&
=  -\frac{1}{2\pi}\sqrt{2}\theta(x^+)\theta(x^-)  \int d^2\vec{r'} \theta(\tau-r') \frac{r'}{r'^2+2(x^{\pm})^2}  f_{IV}(\vec{r}+\vec{r'}).\notag\\&
\end{align}

\underline{Case V: $\delta'(x^{\pm})\theta(x^{\mp})\delta^{(2)}(\vec{r}-\vec{b_i})$ terms}
\vspace{0.2in}

This is a combination of cases $III$ and $IV$ with $\vec{r_i}=\vec{r}-\vec{b_i}$ with $i=1,2$ as in (\ref{r12}). We have

\begin{align}\label{c5}
 G\otimes & \delta'(x'^{\pm})\theta(x'^{\mp}) f_{V}(\vec{r'})\delta^{(2)}(\vec{r}-\vec{b_i})=-
   \int_{-\infty}^{\infty} \frac{ d x'^{\pm}  d x'^{\mp}}{4\pi} \theta(x^{\pm}-x'^{\pm})\theta(x^{\mp}-x'^{\mp}) \delta(x'^{\pm}) \theta(x'^{\mp})  \notag\\&
 \times (-\partial_{x'^{\pm}})   \int d^2\vec{r'} \frac{ \delta \left(\sqrt{2 (x^{\pm}-x'^{\pm})(x^{\mp}-x'^{\mp})}- |\vec{r}-\vec{r'}|\right)}{\frac{1}{\sqrt{2}}\left( (x^{\pm}-x'^{\pm})+(x^{\mp}-x'^{\mp}) \right)}  f_{V}(\vec{r'}) \delta^{(2)}(\vec{r}-\vec{b_i})   \notag\\&
=-\frac{ f_{V}(\vec{b}_i)}{4\pi } \theta(x^{\pm}) \partial_{x^{\pm}} \left \{    \int _{-\infty}^{\infty}  d x'^{\mp} \frac{\theta(x^{\mp}-x'^{\mp}) \theta(x'^{\mp})}{\frac{1}{\sqrt{2}}\left( x^{\pm}+(x^{\mp}-x'^{\mp}) \right)}    
\delta \left(\sqrt{2 x^{\pm}(x^{\mp}-x'^{\mp})}- r_i \right) \right\}  \notag\\&
=-\frac{ f_{V}(\vec{b}_i)}{4\pi } \theta(x^{\pm})  \theta(x^{\mp})\partial_{x^{\pm}} \left \{    \int_{0}^{x^{\mp}} d x'^{\mp} \frac{\sqrt{2}}{\left( x^{\pm}+(x^{\mp}-x'^{\mp}) \right)}    
\delta \left(\sqrt{2 x^{\pm}(x^{\mp}-x'^{\mp})}- r_i \right) \right\}  \notag\\&
=-\frac{1}{2\pi } \sqrt{2} \theta(x^{\pm})  \theta(x^{\mp})\partial_{x^{\pm}} \left \{ \frac{r_i}{r_i^2+2 (x^{\pm})^2} \theta(\tau-r_i) \right\}   f_{V}(\vec{b}_i) \notag\\&
\end{align}
where in the first equality we ignored a term similar to case $II$ (see term (a)) when integrating by parts the $\delta'(x'^{\pm})$ while in the second equality we performed the $\delta(x'^{\pm})$ integration and exchanged $\partial_{x'^{\pm}}$ with $-\partial_{x^{\pm}}$. The rest two steps are obvious.

\chapter{Evaluating the Integral (5.4)}\label{B}


We wish to evaluate the expression (\ref{c6}) by performing the integration on both the light-cone and the transverse plane. We begin by performing the $x^+$ and $x^-$ integrations. Using that $\int dx^-\int dx^- \theta(x^-)=x^- \theta(x^-)$ we find 
\begin{align}\label{A1}
G&\otimes \int dx^-\int dx^- \Big(t_{2,x'^1} \nabla_{\bot}^2 t_{1,x'^1}  +  t_{2,x'^2} \nabla_{\bot}^2 t_{1,x'^2} \Big) \notag\\&
=-\frac{1}{4\pi }\mu^2 \theta(x^+) 
\int_{-\infty}^{\infty}dx'^- x'^- \theta(x'^-) \theta(x^- -x'^-)    \notag\\&
 \hspace{0.8in}\times \int d^2\vec{r'}      \frac{\delta \left(\sqrt{2 x^+(x^--x'^-)}- |\vec{r}-\vec{r'}| \right)} {\frac{1}{\sqrt{2}}\left( x^+ +(x^--x'^-) \right)} \left(  t_{2,x'^1} \nabla_{\bot}^2 t_{1,x'^1}  +  t_{2,x'^2} \nabla_{\bot}^2 t_{1,x'^2}  \right) \notag\\&
 = -\frac{1}{2\pi}\frac{\mu^2}{\sqrt{2}x^+}\theta(x^+) \theta(x^-)  \int d^2\vec{r'}  |\vec{r}-\vec{r'}|\frac{\tau^2-|\vec{r}-\vec{r'}|^2}{|\vec{r}-\vec{r'}|^2+2 (x^+)^2} \theta(\tau-|\vec{r}-\vec{r'}|) \notag\\&
\hspace{2.5in} \times \left(  t_{2,x'^1} \nabla_{\bot}^2 t_{1,x'^1}  +  t_{2,x'^2} \nabla_{\bot}^2 t_{1,x'^2}  \right)
\end{align}
where after the first equality we assume that the $t_{1,2}=\log(k |\vec{r}-\vec{b}_{1,2}|)$ while the $x^{\pm}$ dependence is displayed explicitly and it is integrated out after the second equality. The next step is to perform the transverse integrations. The trick here is to integrate by parts the $\nabla_{\bot}^2t_{1,x^{1,2}}=\partial_{x^{1,2}}(\nabla_{\bot}^2t_1)$ terms. The by parts integration produces two kind of terms: (a) those that do not act on $t_{2,x^{1,2}}$ and (b) those that act on $t_{2,x^{1,2}}$. But the terms of case (b) are proportional to 
\begin{align}\label{nab^2}
\nabla_{\bot}^2 t_2 \nabla_{\bot}^2 t_1 \sim \delta(\vec{r}-\vec{b}_2) \delta(\vec{r}-\vec{b}_1) \sim  \delta(\vec{b}_2-\vec{b}_1)
\end{align}
 which is zero for non zero impact parameter $\vec{b}_2-\vec{b}_1$ while for zero impact parameter it diverges violently; we conclude that an  impact parameter  is necessary (see section \ref{scon}). We now proceed to the remaining terms. Exchanging $\partial_{x'^{1,2}}$ with $-\partial_{x^{1,2}}$ and using (\ref{dlog}) equation (\ref{A1}) gives
\begin{align}\label{A2}
G&\otimes \int dx^-\int dx^- \Big(t_{2,x'^1} \nabla_{\bot}^2 t_{1,x'^1}  +  t_{2,x'^2} \nabla_{\bot}^2 t_{1,x'^2} \Big) \notag\\&
=-\frac{\mu^2}{\sqrt{2}x^+} \theta(x^+) \theta(x^-)\Bigg \{ \partial_{x^1} \int d^2\vec{r'}  |\vec{r}-\vec{r'}|\frac{\tau^2-|\vec{r}-\vec{r'}|^2}{|\vec{r}-\vec{r'}|^2+2 (x^+)^2} \theta(\tau-|\vec{r}-\vec{r'}|) \notag\\&
\hspace{2.2in} \times  \frac{x'^1-b_{21}}{|\vec{r'}-\vec{b}_2|^2}  \delta(\vec{r'}-\vec{b}_1) +\big(1 \leftrightarrow 2 \big) \Bigg \} \notag\\&
=-\frac{\mu^2}{\sqrt{2}x^+} \theta(x^+) \theta(x^-) \left \{\frac{b_{11}-b_{21}}{|\vec{b}_2-\vec{b}_1|} \partial_{x^1}  \left(\theta(\tau-r_1) r_1 \frac{\tau^2-r_1^2}{r_1^2+2 (x^+)^2} \right)+ \big(1 \leftrightarrow 2 \big) \right \} \notag\\&
=-\frac{\mu^2}{\sqrt{2}x^+} \theta(x^+) \theta(x^-) \left \{\frac{b_{11}-b_{21}}{|\vec{b}_2-\vec{b}_1|^2}\theta(\tau-r_1)  \partial_{x^1}  \left(r_1 \frac{\tau^2-r_1^2}{r_1^2+2 (x^+)^2} \right)+ \big(1 \leftrightarrow 2 \big) \right \}
\end{align}
where $\vec{b}_{1,2}$ are given by (\ref{b}).


\chapter{Evaluating the Integral (5.8)}\label{C}

We wish to calculate the integral (\ref{Jin}) that we encountered in section \ref{TP}. We have
\begin{align} \label{J1}
{\cal J} (r_1,r_2,\tau)=\frac{1}{2 \pi \tau}\int_{0}^{\infty}\int_{0}^{2\pi}r'dr'd\phi' \delta(\tau-r') \log(k |\vec{r'}+\vec{r_1}|) \log(k |\vec{r'}+\vec{r_2}|).
\end{align}
The quantities $r_{1,2}$ are given by (\ref{r12}). The trick here is to expand the logarithms in their Fourier space: $\log(kr)=-\int \frac{d^2q}{2\pi}\frac{ e^{i \vec{q} \hspace{0.02in} \vec{r}}}{q^2}$ with $k$ serving as an ultraviolet cutoff.
Expanding both logarithms and performing the angular integration one obtains
\begin{align}\label{J2}
{\cal J}=\frac{1}{\tau} \int_{0}^{\infty}dr' r'\delta(\tau-r') \left \{ \int\frac{d^2q d^2l}{(2\pi)^2} \frac{e^{i \vec{q} \hspace{0.02in} \vec{r_1}+i \vec{l} \hspace{0.02in} \vec{r_2} }} {q^2 l^2} J_0 (r'|\vec{q}+\vec{l}|) \right\}.
\end{align}
Performing the trivial radial integration we find that${\cal J}$ is now defined by
\begin{align}\label{J3}
{\cal J} \equiv \int\frac{d^2q d^2l}{(2\pi)^2} \frac{e^{i \vec{q} \hspace{0.02in} \vec{r_1}+i \vec{l} \hspace{0.02in} \vec{r_2} }} {q^2 l^2} J_0 (\tau|\vec{q}+\vec{l}|)  
\end{align}
Next we perform the $l$ and $q$ integrations. These integrals have been calculated in \cite{Kovchegov:1997ke}; we summarize the procedure: In order to perform these integrations one has to expand $ J_0 (\tau|\vec{q}+\vec{l}|)$ in an infinite sum of products of the form $ J_n (\tau|\vec{q}|) J_n (\tau|\vec{l}|)$ with $n$ an integer and do the angular integrals (of q and l) first. This factors out the radial integrations over $l$ and $q$ into two independent integrals. Then one has to perform these integrations and finally sum over $n$. The final result reads

\begin{subequations}\label{Ja}
\begin{align}
{\cal J}(\tau,r_1,r_2)& \equiv \ln(\xi_> k)\ln(\eta_> k)+\frac{1}{4}\left[ Li_2\left( e^{i\alpha}\frac{\xi_<\eta_<}{\xi_>\eta_>} \right) + Li_2\left( e^{-i\alpha}\frac{\xi_<\eta_<}{\xi_>\eta_>} \right)\right], \label{JJ}\\
\xi_{>(<)}&=max(min)(r_1,\tau) \hspace{0.2in} \eta_{>(<)}=max(min)(r_2,\tau), \label{ke}\\
& \hspace{1.2in} (\vec{r_1}).(\vec{r_2})=\cos(\alpha) r_1 r_2. \label{a} 
\end{align}
\end{subequations}
So here $\alpha$ is the angle between $\vec{r}_1$ and $\vec{r}_2$ and $Li_2$ is the dilogarithm function while ${\cal J}$ is real as it should. Equation (\ref{Ja}) implies that ${\cal J}$ depends from the ordering of $r_1$, $r_2$ and $\tau$. There are in principle six distinct ways to order them. However it turns out that  the cases $r_{1,2}>\tau$ and $r_{1,2}<\tau$ are interdependent from the relative ordering of $r_1$ and $r_2$. This degeneracy reduces the possible cases to four which we organize by introducing table \ref{ta1} \footnote{${\cal J}_3$ for instance means ${\cal J}(\xi_>\eta_>=\tau r_2,\xi_<\eta_<=r'r_1)$ with ${\cal J}(\xi_>\eta_>,\xi_<\eta_<)$ given from (\ref{Ja}). } that in turn help us to write a (unified) formula for ${\cal J}$
\begin{align}\label{Jap}
{\cal J} (r_1,r_2,\tau)& = \theta(r_1-\tau)\theta(r_2-\tau){\cal J}_1 (r_1,r_2,\tau)+ \theta(\tau-r_2)\theta(r_1-\tau){\cal J}_2 (r_1,r_2,\tau) \notag\\&
+ \theta(\tau-r_1)\theta(r_2-\tau){\cal J}_3 (r_1,r_2,\tau)+\theta(\tau-r_1)\theta(\tau-r_1){\cal J}_4 (r_1,r_2,\tau).
\end{align}
\begin{table}
\caption{We defined $\vec{r}_{1,2}=\vec{r}-\vec{b}_{1,2}$}
\centering
\begin{tabular}{c|cccc|c|c}
\hline\hline
cases & $\xi_>$ & $\eta_>$ & $\xi_>\eta_>$ & $\xi_<\eta_<$ & ${\cal J}_i$ & region (see figure \ref{re})\\
\hline\hline
1 & $r_1$ & $r_2$ & $r_1r_2$ & $\tau^2$ & ${\cal J}_1$ & I\\
2 & $r_1$ & $\tau$ & $\tau r_1$ & $\tau r_2$ & ${\cal J}_2$ & II$^{\prime}$\\
3 & $\tau$ & $r_2$ & $\tau r_2$ & $\tau r_1$ & ${\cal J}_3$ &II\\
4 & $\tau$ & $\tau$ & $\tau^2$ & $r_1r_2$ & $ {\cal J}_4$ &III\\
\hline\hline
\end{tabular}
\label{ta1}
\end{table}

\clearpage 


\end{document}